\documentclass[prb,twocolumn,amsmath,amssymb,aps,floatfix,superscriptaddress]{revtex4-2} 
\usepackage{graphicx}
\usepackage{amssymb}
\usepackage{amsmath}
\usepackage{physics}
\usepackage{xcolor}
\usepackage{lipsum}
\usepackage{upgreek}
\usepackage{bm}
\usepackage{hyperref}
\usepackage{mathrsfs}
\newcommand{\rmm}[1]{{\textcolor{black}{#1}}}
\newcommand{\rai}[1]{{\textcolor{black}{#1}}}

\newcommand{\vc}[1]{\vec{#1}}
\usepackage{comment}
\usepackage{soul}
\usepackage[normalem]{ulem}

\begin{document}

\preprint{APS/123-QED}

\title{\textit{In silico} design of magnonic lasing in constricted waveguides}

\author{Jean F. O. da Silva}
\affiliation{COMMIT, Department of Physics, University of Antwerp, Groenenborgerlaan 171, B-2020 Antwerp, Belgium\\}
\author{Rai M. Menezes}%
\affiliation{COMMIT, Department of Physics, University of Antwerp, Groenenborgerlaan 171, B-2020 Antwerp, Belgium\\}
\affiliation{Departamento de Física, Universidade Federal de Pernambuco, Cidade Universitária, 50670-901 Recife-PE, Brazil}
\author{Milorad V. Milo\v{s}evi\'c}%
\email{milorad.milosevic@uantwerpen.be}
\affiliation{COMMIT, Department of Physics, University of Antwerp, Groenenborgerlaan 171, B-2020 Antwerp, Belgium\\}

\begin{abstract}
Analogue black-hole systems have been proposed in various physical platforms, including magnetic materials, offering rich physics and promising applications such as wave lasing. However, their practical realization and characterization remain largely unexplored. Here, we present an \textit{in silico} study of magnonic black-hole phenomena in constricted ferromagnetic waveguides driven by spin-polarized currents. Using micromagnetic simulations with a two-dimensional Poisson solver to obtain realistic current-density profiles, we demonstrate the formation of a double-hole cavity bounded by analogue event horizons, enabling resonant spin-wave amplification. We characterize the resonances as a function of magnetic and geometric parameters and identify the corresponding spin-wave modes. We further show that gradual tapers enhance transmission by suppressing spin-wave reflections, while interfacial Dzyaloshinskii--Moriya interaction can mimic the current-induced Doppler shift, substantially reducing the critical current density required for applications and, consequently, Joule heating. Finally, we demonstrate a spin-wave laser in which thermally excited spin waves undergo selective amplification and coherent emission at well-defined resonance frequencies. These results provide design principles for magnonic analogues of gravitational systems and point toward their potential for advanced spintronic applications.

\end{abstract}

\date{\today}

\maketitle

\section{\label{Intro}Introduction}

Collective spin excitations in magnetic materials give rise to magnons, quasiparticles associated with correlated spin precessions that propagate as spin waves (SWs)~\cite{rezende2020fundamentals}. These excitations form the basis of magnonics, enabling the transport and manipulation of information through wave-like phenomena. Owing to their high operational frequencies in the GHz--THz range~\cite{jin2018ramanTHz,menezes2022tailoring_THz}, controllability via path manipulation and magnetic fields~\cite{menezes2024towards,Spin_torch_paper}, and their ability to exhibit interference and phase-dependent effects~\cite{Mach-Zehnder_SW,Mach-Zehnder_SW_logic}, spin waves offer a promising platform for wave-based logic and signal processing\rmm{~\cite{flebus20242024}. They have also been proposed as a potential platform for neuromorphic computing~\cite{menezes2024towards,papp2021nanoscale}}. Moreover, SWs propagate without charge carriers, thereby eliminating Joule heating and offering exceptional energy efficiency for prospective devices.

\rmm{On the other hand, the application of an electric current in a thin magnetic film generates a Spin-Transfer Torque (STT), which interacts with the magnetization dynamics and gives rise to a Doppler shift in the spin waves, characterized by an effective “spin-drift” velocity $v_s$, as demonstrated in both theoretical and experimental works~\cite{Zhang_Li_adiabatic,Zhang_Li_non_adiabatic,SW_doppler_exp}. As in other wave phenomena in physics, the Doppler shift is limited by a critical velocity, namely the spin-wave velocity $c$, at which the wave system undergoes an analogous transition from a “subsonic” regime ($v_s<c$) to a “supersonic” regime ($v_s>c$).}

\rmm{Recent theoretical works have demonstrated that, in spatially engineered current distributions (e.g., in a constricted waveguide), where localized regions of the sample satisfy $v_s>c$, the spin-wave (SW) system behaves as an analogue black-hole system~\cite{MBH_molina,SWLasing}. Specifically, such a system is termed a double-hole cavity (DHC) because it effectively hosts a magnonic analogue of a black hole (MBH) on one side and a magnonic white hole (MWH) on the other. These structures emerge when the dispersion relation transitions between positive and negative frequencies, and vice versa, as the drift velocity crosses the critical value. Crucially, the velocity transition regions correspond to event horizons, a concept borrowed from analogue gravity systems~\cite{barcelo2019analogue}.}

\rmm{The transmission of SWs through DHCs in constricted waveguides has recently been investigated theoretically, revealing intriguing phenomena such as resonant amplification of spin waves and SW lasing~\cite{wang2024supermirrors,nakayama2024resonant_christmas_eve}. However, the pioneering theoretical studies in this context~\cite{MBH_molina,SWLasing} lacked a more realistic physical framework. For instance, the analytical calculations assumed step-like spatial distributions of the current density, which do not accurately represent real experimental systems. In addition, the current densities considered in previous simulations were exceedingly high for practical applications. These limitations highlight the need for more advanced engineering of magnonic gravity analogues, aimed at reducing the required current densities and achieving conditions compatible with experimental realization. Addressing these challenges requires a deeper understanding of magnonic black-hole phenomena, as well as the identification of the key material and geometrical parameters governing the formation and control of gravity analogues on a chip.}

In this work, we present a comprehensive theoretical and computational study of double-hole cavity (DHC) phenomena in constricted magnetic waveguides subjected to electric currents. Figure~\ref{Setup_config} illustrates the considered system: a metallic ferromagnetic waveguide of width $W$ containing a bow-tie constriction at its center. In contrast to previous studies, we explicitly account for realistic spatial distributions of the current density and other relevant physical effects, with the aim of characterizing and optimizing the predicted resonant amplification for spin-wave lasing applications. We begin by investigating the formation of the magnonic cavity through calculations and analysis of the system dispersion relation, followed by the determination of the critical current required to achieve gravity-analogue regimes in realistic samples. We then reproduce the resonant amplification previously predicted for idealized step-like current profiles and characterize its dependence on the DHC size and spin-wave frequency. Subsequently, we extend the analysis to realistic current distributions by implementing a Poisson solver to obtain the two-dimensional current profile in the constricted waveguide. The resonant amplification is then investigated as a function of the geometrical parameters governing the current distribution, as well as key magnetic parameters, including the perpendicular magnetic anisotropy (PMA) and the interfacial Dzyaloshinskii--Moriya interaction (iDMI). We demonstrate that the iDMI, induced for example by a substrate or an adjacent heavy-metal layer, can effectively mimic the role of the applied current in the magnonic cavity, thereby reducing both the threshold current and the associated Joule heating. This significantly facilitates the experimental realization of magnonic gravity analogues under more accessible conditions.
Finally, we demonstrate the realization of spin-wave lasing by incorporating thermal excitations into the constricted waveguide. We show that the double-hole cavity can selectively amplify thermally excited SWs through resonant amplification, a mechanism predicted analytically in the literature~\cite{SWLasing} but not previously realized. This leads to the generation and coherent emission of spin waves at well-defined resonance frequencies.

\rmm{The paper is organized as follows. In Sec.~\ref{Theory}, we present the theoretical framework describing spin-wave dispersion in analogue black-hole systems, together with details of the micromagnetic simulations employed in this work. In Sec.~\ref{sec.results}A, we characterize the considered system by calculating the spin-wave dispersion of the material and determining the critical current required to induce the analogue black-hole regime. In Sec.~\ref{sec.results}B, we demonstrate and characterize the resonant amplification in a DHC for both step-like and Gaussian-like current profiles. In Sec.~\ref{sec.results}C, we investigate resonant amplification in constricted waveguides with realistic current distributions obtained from a Poisson solver, analyzing its dependence on the relevant geometrical parameters.} \rai{In Sec.~\ref{sec.results}D, we investigate the influence of the PMA on resonant amplification. In Sec.~\ref{sec.results}E, we show that the iDMI can effectively mimic the role of the applied current, providing an alternative route to reduce the current density required for practical applications. In Sec.~\ref{sec.results}F, we demonstrate the realization of SW lasing through the resonant amplification of thermally excited spin waves generated in the heated constriction, resulting in coherent spin-wave emission. Finally, our main findings and conclusions are summarized in Sec.~\ref{sec:End}.}

\section{\label{Theory}Theoretical framework}

\subsection{\label{sec:LLG}Spin-wave dispersion relation}
\begin{figure}[t]
\includegraphics[width=\linewidth]{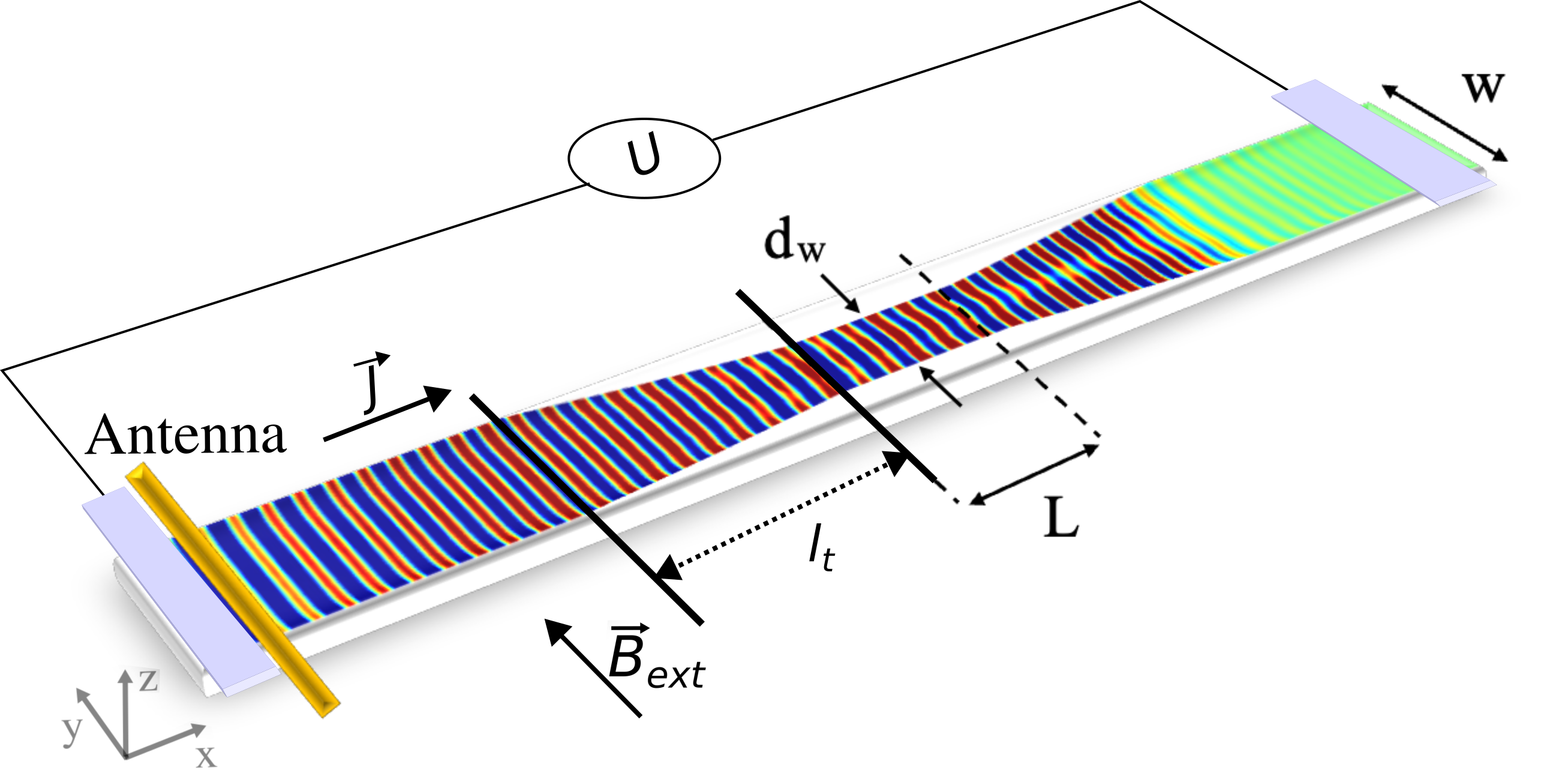}
\caption{\label{Setup_config} Schematic illustration of the studied system. A metallic ferromagnetic waveguide of width $W$, with a bow-tie constriction at its center. SWs are generated by an antenna on the left-hand side and propagate along the $+\hat{x}$-direction. An external bias magnetic field, $\vec{B}_{ext}$, is applied in the $+\hat{y}$-direction. The constriction width, $d_w$, length of the constriction, $L$, and taper length, $l_t$, are varied to study the effects of spatial current gradients on SW propagation.}
\end{figure}

Regarding the micromagnetic model, we describe the magnetization of a thin ferromagnetic film as $\vec{M}(\bf r)$ = $M_S\vec{m}(\bf r)$, where $\vec{m}(\bf r)$ is a normalized vector field, and $M_S$ is the saturation magnetization.
The magnetization dynamics is ruled by the Landau-Lifshitz-Gilbert (LLG) equation \cite{gilbert2004phenomenological}
\begin{equation}
    \dot{\vec{m}} = -\gamma\vec{m}\times\vec{H}_{eff} + \alpha\vec{m}\times\dot{\vec{m}} + \vec{\tau}_{STT} \label{LLG_equation}.
\end{equation} 
Here, $\gamma$ is the gyromagnetic ratio; $\vec{H}_{eff}$ is the effective field, which can be derived from $\vec{H}_{eff}$ = -$\delta E/\delta\vec{M}$, with $E$ the magnetic energy functional; $\alpha$ is the Gilbert damping factor and $\vec{\tau}_{STT}$ accounts for current-induced spin-transfer-torques (STT)~\cite{RALPH20081190, Zhang_Li_adiabatic, Zhang_Li_non_adiabatic}. For the energy functional, we take into
account the contribution of exchange interaction, \rai{magneto-crystalline anisotropy} and  Zeeman energy due to an applied bias magnetic field:
\begin{equation}\label{eq.Energy}
    \rai{E[\vc{m}] = \int \left[ A(\nabla\vc{m})^2 -K_z m_z^2 -M_s \vc{B}_{ext} \cdot \vc{m} \right] d^3\bf{r},}
\end{equation}
where $A$ is the exchange stiffness, \rai{$K_z$ is the uniaxial anisotropy constant,} and $B_{ext}$ is the bias field, which in this work we consider to be applied along the $y$-direction [cf. Fig.~\ref{Setup_config}]. The effects of other relevant magnetic properties, such as dipolar interactions, are discussed in the \rai{supplemental material}~\cite{SM}.

When considering current-induced STT, the Zhang-Li torque $\tau_{STT}$ is to be added to the LLG equation \cite{Zhang_Li_adiabatic,Zhang_Li_non_adiabatic}. Considering small variations of magnetization from the ground state, where $\vec{m}$ is saturated in the $y$-direction, such that $\nabla \vec{m} \approx (\frac{\delta m}{\delta x}, 0, \frac{\delta m}{\delta z})$ one can rewrite the LLG equation accounting for STT as \cite{MBH_molina, SWLasing}:
\begin{equation}
    (\partial_t + \vec{v}_s \cdot \nabla)\vec{m} = -\gamma\vec{m} \times \vec{H}_{eff} + \alpha \vec{m}\times (\partial_t + \frac{\beta}{\alpha}\vec{v}_s\cdot \nabla)\vec{m} \label{equation_doornenbal},
\end{equation}
where $\beta$ is the non-adiabatic STT constant and $\vec{v}_s = \frac{P \mu_B}{e M_S (1+\beta^2)}\vec{J}$, with $P$ the spin polarization of the current, $\vec{J}$ the current density, $\mu_B$ the Bohr magneton, and $e$ the electron charge. 

Considering a linearized spin-wave solution for the above LLG equation, with $\alpha$ and $\beta$ small, one can obtain the following dispersion relation:

\begin{equation}
   \omega - k{v_s}= \pm\sqrt{\left[\frac{2\gamma A}{M_S}k^2+ \omega_0 \right]^2 } \label{disp_STT},
\end{equation}
where $\omega$ and $k$ are the SW frequency and wavenumber, respectively. Here $\omega_0 = \gamma B_\text{ext}$ is the frequency for $k=0$. The derivation of Eq.~\eqref{disp_STT} can be found in the supplemental material~\cite{SM}. 

\subsection{\label{sec:conditions_MBH} Analogue black-hole horizons}

\begin{figure}[]
\includegraphics[width=0.8\linewidth]{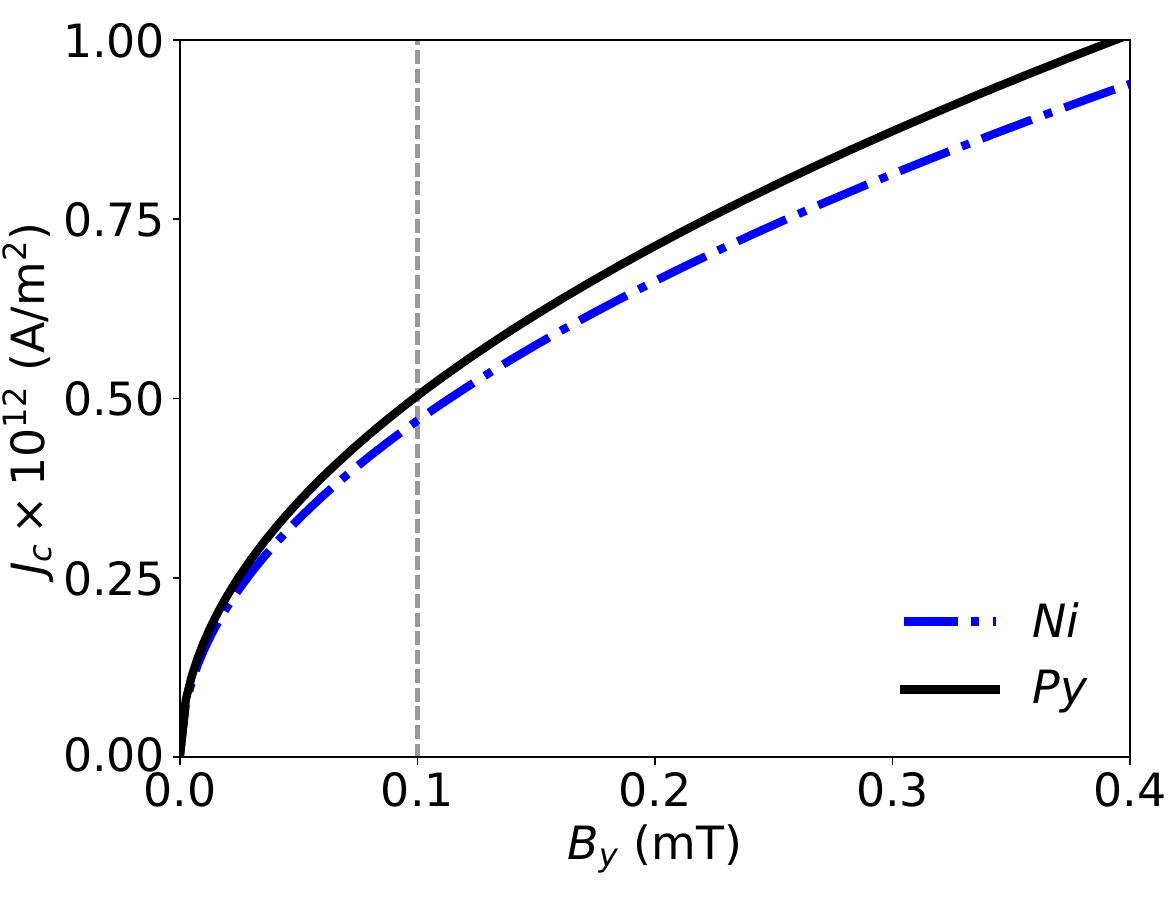}
\caption{\label{Jc_vs_By} Critical current as a function of the external field $B_{y}$ for permalloy and nickel thin films (see parameters in Sec.~\ref{Methods}). The gray dashed line corresponds to $B_y$ = 0.1 mT, the bias field considered in Sec. \ref{sec.results}. }
\end{figure}
From Eq.~\eqref{disp_STT}, one notices that by applying a spin-polarized electric current, it induces a frequency shift in the SW analogous to a Doppler shift, with an effective spin-drift velocity $v_s$. This behavior of the wave dispersion resembles what is observed in analogue black hole systems in the literature \cite{mayoral2011acoustic,barcelo2019analogue,butera2017blackhole,rousseaux2020classical,recati2009bogoliubov,Leonhardt2007_BHL_revisited}. When $v_s$ exceeds the critical spin-wave velocity $c$, the system undergoes a transition analogous to the ``subsonic" to ``supersonic" transition. The critical value of $v_s$ is obtained when the dispersion relation in Eq.~\eqref{disp_STT} yields negative frequency solutions.  

Analogue black-hole horizons are therefore observed in the spin-wave system when the current density $J$ varies along the magnetic film, such as in constricted waveguides. These horizons delineate regions where $v_s < c$ from regions where $v_s > c$.  

The critical current that defines the black-hole horizon is calculated from Eq.~\eqref{disp_STT}. Therefore, one finds:  
\begin{equation}
     J_c = \frac{2e\gamma}{P\mu_B}\sqrt{2M_S A B_{ext}}. \label{J_c}
\end{equation}
From this equation, it is clear that selecting a material with a low exchange stiffness constant $A$ can reduce the critical current density $J_c$, thereby facilitating the creation of the MBH. Figure~\ref{Jc_vs_By} shows the critical current as a function of the applied field, calculated for conducting and ferromagnetic thin films of nickel (Ni) and permalloy (Ni$_{80}$Fe$_{20}$, Py), both often used in spintronics literature~\cite{rezende2020fundamentals,Ni_Py_parameters,exchange_stiffness_parameters,Ni_polarization,SW_doppler_exp}. They were chosen due to their low magnetization saturation and high polarization, i.e. $M_{S,Py} = 750$~kA/m, $M_{S,Ni} = 490$~kA/m \cite{Ni_Py_parameters,SW_doppler_exp}.
Although Ni exhibits a lower critical current compared to permalloy (i.e., $J_{c,\mathrm{Ni}} \big/ J_{c,\mathrm{Py}} \approx 93\%$), the latter was preferred in this work, as Py thin films are more used in experiment and better documented in the literature.

\subsection{\label{Methods}Micromagnetic simulations}

To simulate SW propagation, we employed the Mumax$^3$ package~\cite{Mumax_original2014}. The simulations were performed for a permalloy strip of length 8000 $\,\mathrm{nm}$, nominal width $w=800\,\mathrm{nm}$, and thickness $20\,\mathrm{nm}$, with $ M_s = 750\,\mathrm{kA/m} $, $\gamma / 2\pi = 29\,\mathrm{GHz/T}$, and exchange stiffness $ A = 10\,\mathrm{pJ/m} $. We use micromagnetic cells of size \( 4 \times 4 \times 4\,\text{nm}^3 \). For all simulations, the bias magnetic field was set to \( B_{\mathrm{ext}} = 0.1\,\text{mT} \) and applied along the \( \hat{y} \) direction. \rai{Initially, the anisotropy constant is set to $K_z=0$, and its contribution is explored later in a dedicated section.} To excite SWs in the system, we positioned an antenna at the left-hand end of the sample, where an oscillating magnetic field is applied [see Fig.~\ref{Setup_config}].
 
In Section~\ref{MBH_real}, a bow-tie constriction is introduced at the center of the sample by locally reducing the waveguide width to $\mathrm{d_w}$, with a taper length $\mathrm{l_t}$, as illustrated in Fig.~\ref{Setup_config}. To accurately compute the interaction of SWs with non-uniform current distributions in this constricted geometry, we implemented a Poisson solver in the micromagnetic simulations~\cite{menezes2024towards}, via a custom module added to the Mumax$^3$ simulation package. In this approach, for a set of contact micromagnetic cells where a bias voltage \( U \) is fixed, the generalized Poisson equation \( \nabla \cdot (\sigma[\vec{m}] \nabla \Phi) = 0 \) is solved using a finite-difference gradient method. The solution yields the current density \( \vec{j} = -\sigma[\vec{m}] \nabla \Phi \), where \( \sigma[\vec{m}] \) is the conductivity tensor and \( \Phi \) is the electric potential, which satisfies the Poisson equation with the boundary condition \( \Phi|_{\text{contact,L}} = -\frac{U}{2} \), \( \Phi|_{\text{contact,R}} = \frac{U}{2} \). For further details on the method, we refer the reader to Ref.~\cite{menezes2024towards}.

\section{\label{sec.results}Results and discussion}

\subsection{\label{sec:Doppler_shift}Spin-wave dispersion under applied current}

\begin{figure}[t!]
\includegraphics[width=\linewidth]{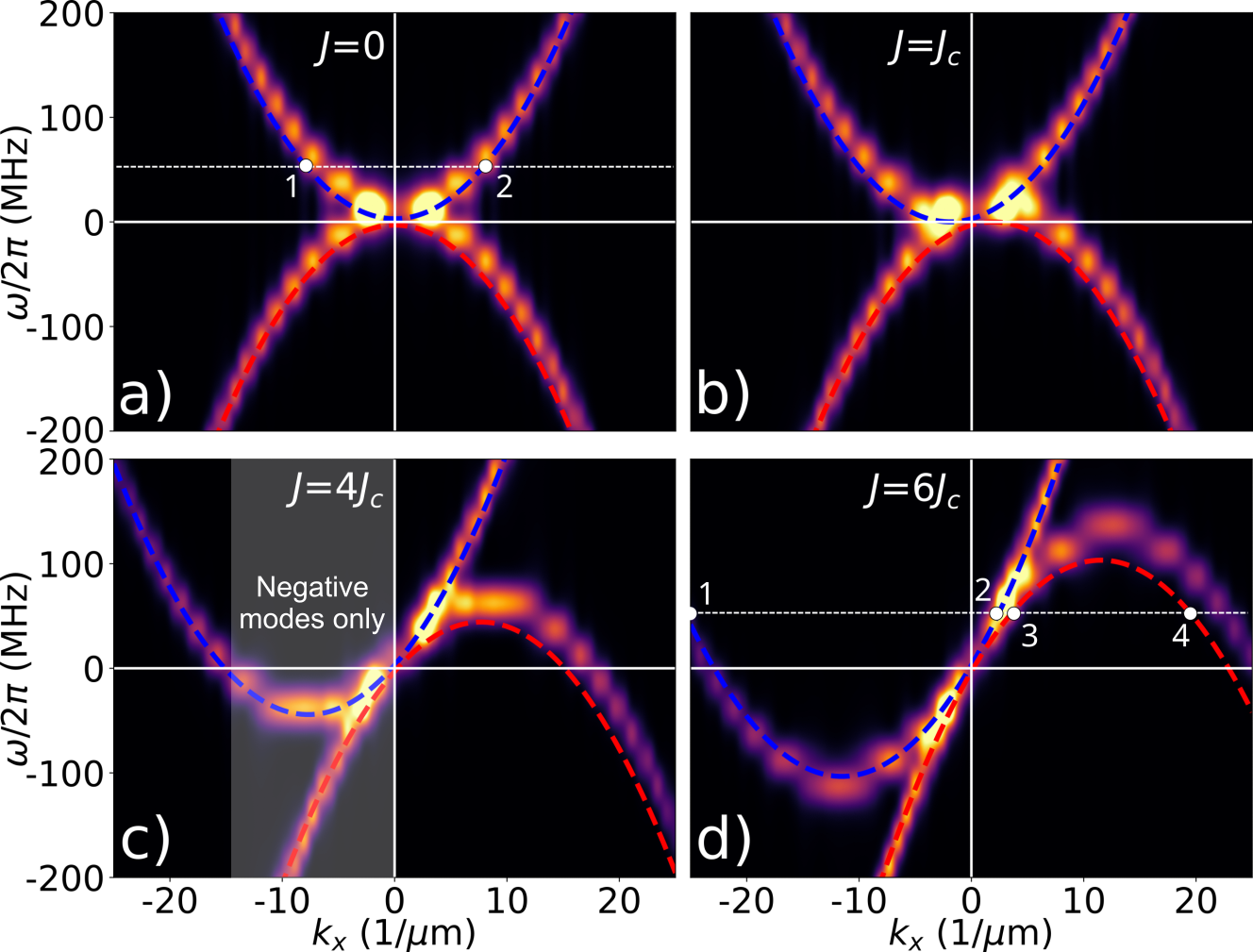}
\caption{\label{Disp_rel_J1} Dispersion relation for the unconstricted permalloy strip of width $\mathrm{w}=0.8$~$\upmu$m and applied current (a) $J = 0$, (b) $J = J_c$, (c) $J = 4J_c$, and (d) $J = 6J_c$, where $J_c = 5.03 \times 10^{11}\,\text{A/m}^2$ is the critical current. The red and blue dashed lines correspond to the $+\omega$ and $-\omega$ solutions of Eq.~\eqref{disp_STT}, respectively. The white dots in (a) and (d) indicate the possible propagating modes for a chosen frequency $\omega$. The shaded region in (c) highlights the wave-vector range where only negative-energy modes are allowed.}
\end{figure}

\rai{To compute the SW dispersion relations, we consider a permalloy strip with the dimensions and material parameters described in Sec.~\ref{Methods}, without the constriction. SWs are excited at the center of the waveguide using an antenna that generates an oscillating magnetic field, $\vec{B}_{\mathrm{ant}} = B_0\,\mathrm{sinc}(2\pi f_{\max} t)\,\hat{z}$, where $B_0 = 5 \,\mathrm{\upmu T}$ is the field amplitude and $f_{\max}=200~\mathrm{MHz}$ is the maximum excitation frequency. Periodic boundary conditions are applied perpendicular to the propagation direction to ensure a continuous representation of the SW dynamics. Figure~\ref{Disp_rel_J1} shows the dispersion relations obtained for different values of the applied current, with the analytical solutions of Eq.~(\ref{disp_STT}) shown as dashed lines for comparison. For the considered parameters, Eq.~(\ref{J_c}) yields a critical current density of $J_c = 5.03 \times 10^{11}~\mathrm{A/m^2}$.} 

Figure~\ref{Disp_rel_J1}(a) shows the dispersion relation in the absence of an applied current, where the characteristic parabolic shape is evident. Typically, for \( J < J_c \), the dispersion exhibits a gap, and one requires \( \omega \geq \omega_{\text{min}} \) for scattering solutions to exist (for \( J = 0 \), \( \omega_{\text{min}} = \omega_0 \)). In this regime, there are two propagating modes with real \( k \) for any frequency above \( \omega_{\text{min}} \), as indicated by the white dots in Fig.~\ref{Disp_rel_J1}(a). 

On the other hand, when \( J > J_c \) [see Fig.~\ref{Disp_rel_J1}(c,d)], there exists a frequency range with \( \omega < \omega_{\text{max}} \) in which four real wave vectors \( k \) satisfy the dispersion relation, as shown in Fig.~\ref{Disp_rel_J1}(d). For \( \omega > \omega_{\text{max}} \), two of the \( k \) solutions remain real, while the other two become imaginary. Moreover, when \( J > J_c \), there exists a range of wave vectors \( k \) for which only modes with negative energy and frequency are possible, as indicated, for example, by the shaded region in Fig.~\ref{Disp_rel_J1}(c). 

\rai{As discussed in Ref.~\cite{SWLasing}, and as will be shown in the following sections, when the current density varies along the waveguide (for instance, due to a narrowing at a constriction), a region of elevated current density, referred to as a cavity, can form. Within this cavity, SWs are reflected back and forth, allowing spin waves with positive and negative wave vectors, $k$, to interact repeatedly. If the current density within the cavity exceeds \( J_c \), coupling between negative-energy and positive-energy spin waves occurs, potentially leading to resonance~\cite{SWLasing,nakayama2024resonant_christmas_eve}. Such a resonance effect can enhance both the transmission and reflection amplitudes of spin waves scattered within the cavity when the resonance conditions are met. This mechanism is a key ingredient in the conceptualization of SW lasing~\cite{SWLasing}.}


\subsection{\label{sec:DHC_resonance_sharp} Resonant amplification in a DHC}

\subsubsection{DHC with a step-like current profile}\label{sec.step-like}

\begin{figure}[t!]
\includegraphics[width=\linewidth]{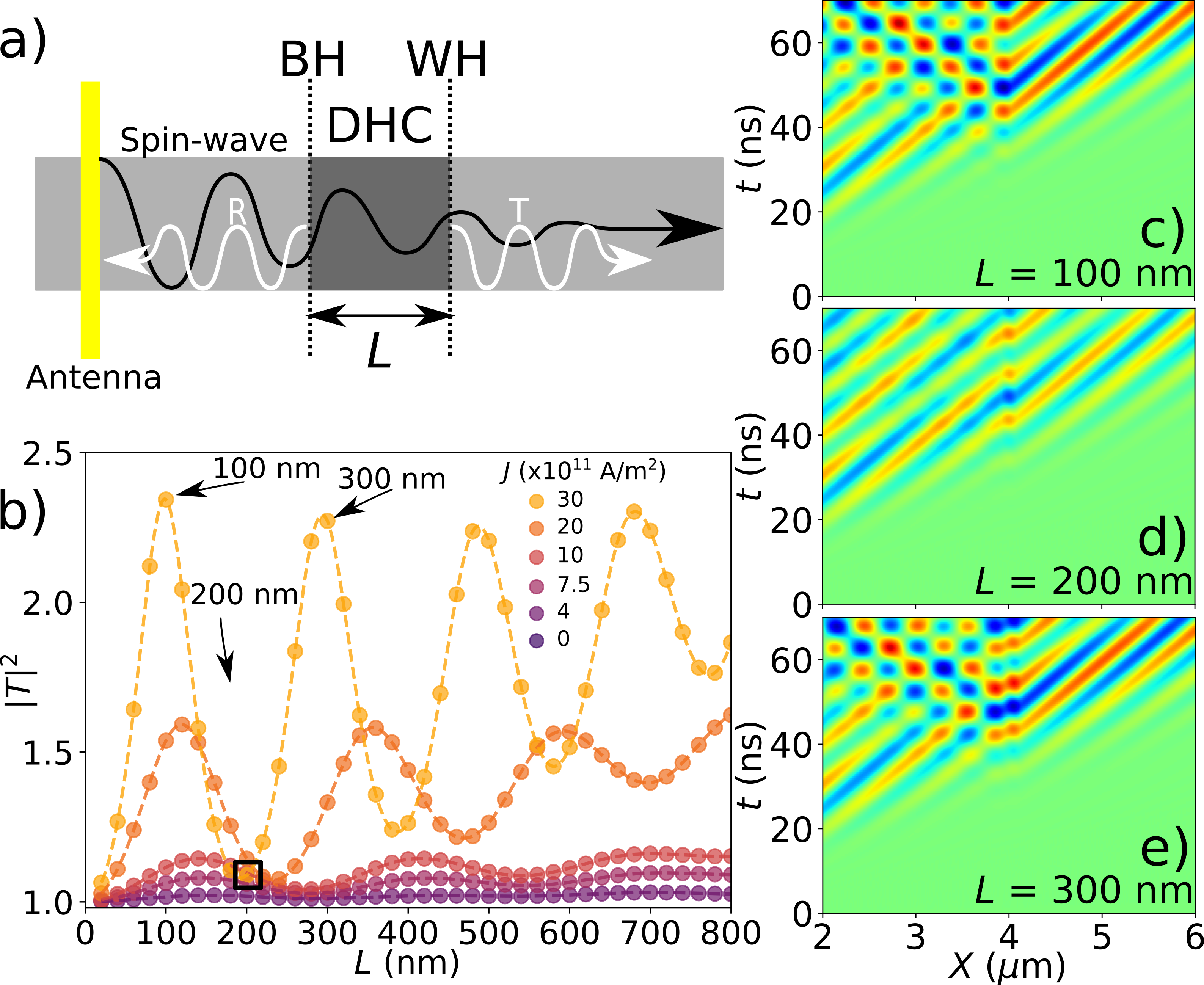}
\caption{\label{DHC_resonance_SW} (a) Setup for spin-wave propagation through a step-like current profile, with the damped wave and horizons indicated. Once the SW enters the DHC, it splits into reflected (R) and transmitted (T) waves. Inside the DHC, SWs can reflect back and forth and may satisfy a resonance condition. 
(b) SW transmission amplitude as a function of cavity length $L$ for different values of current density inside the DHC. The arrows indicate the reference cavity lengths for $J = 30\times10^{11}$ A/m$^2$, shown in panels (c–e). The signal was normalized by the transmission amplitude for $J = 0$. (d–f) Space–time diagrams for the reference points indicated in (b): (d) the resonance peak at $L = 100$ nm, (e) the off-resonance condition at $L = 200$ nm, and (f) the second resonance peak at $L = 300$ nm. The center of the DHC is located in the center of the sample (at $x=4$~$\mu$m).
}
\end{figure}

As discussed in the literature~\cite{nakayama2024resonant_christmas_eve,wang2024supermirrors} and in the Introduction, a DHC for SWs is formed when the current density in the waveguide locally exceeds $J_c$. In this situation, the region where $J > J_c$ (i.e., inside the cavity) is delineated from the regions where $J < J_c$ by the MBH and MWH horizons. To simulate this scenario, we consider a permalloy waveguide in which a current density $J > J_c$ is applied locally to a central section of length $L$, while $J = 0$ elsewhere, resulting in a step-like current profile. \rai{A SW with a fixed frequency $f=\omega/2\pi = 100$~MHz is generated by the antenna field, \( \vec{B}_{\mathrm{ant}} = B_0\,\mathrm{sin}(2\pi f t)\,\hat{z} \), located on the left-hand side of the sample and propagates along the $+\hat{x}$ direction toward the DHC, as illustrated in Fig.~\ref{DHC_resonance_SW}(a). }

Figure~\ref{DHC_resonance_SW}(b) shows the transmitted SW amplitude as a function of the DHC length $L$ for different values of the current density inside the DHC. The signal was calculated immediately after the cavity $L$ in a region of $300 \,\mathrm{nm}$ \rai{length} and it was normalized by the transmission amplitude in the absence of any current ($J = 0$) in the same region. Crucially, transmission peaks characteristic of resonant behavior~\cite{SWLasing,nakayama2024resonant_christmas_eve} emerge for the two highest values of $J$ considered in Fig.~\ref{DHC_resonance_SW}(b), with minor oscillations also observed for the other current density values, except $J = 0 $. 

Figures~\ref{DHC_resonance_SW}(c–e) show space–time diagrams of the SW across the waveguide for selected values of $L$, calculated for the case $J = 30\times10^{11}$~A/m$^2 \approx 6J_c$, with the center of the DHC located at $x = 4~\mu$m. Three reference values of $L$ are shown, corresponding to: (\textit{i}) the resonance peak at $L = 100$~nm [Fig.~\ref{DHC_resonance_SW}(c)]; (\textit{ii}) the off-resonance condition at $L = 200$~nm [Fig.~\ref{DHC_resonance_SW}(d)]; and (\textit{iii}) the second resonance peak at $L = 300$~nm [Fig.~\ref{DHC_resonance_SW}(e)]. The resonance scenarios are characterized by standing waves inside the cavity and by an enhancement of both the transmitted and reflected wave amplitudes.

\begin{figure*}[t]
\includegraphics[width=\linewidth]{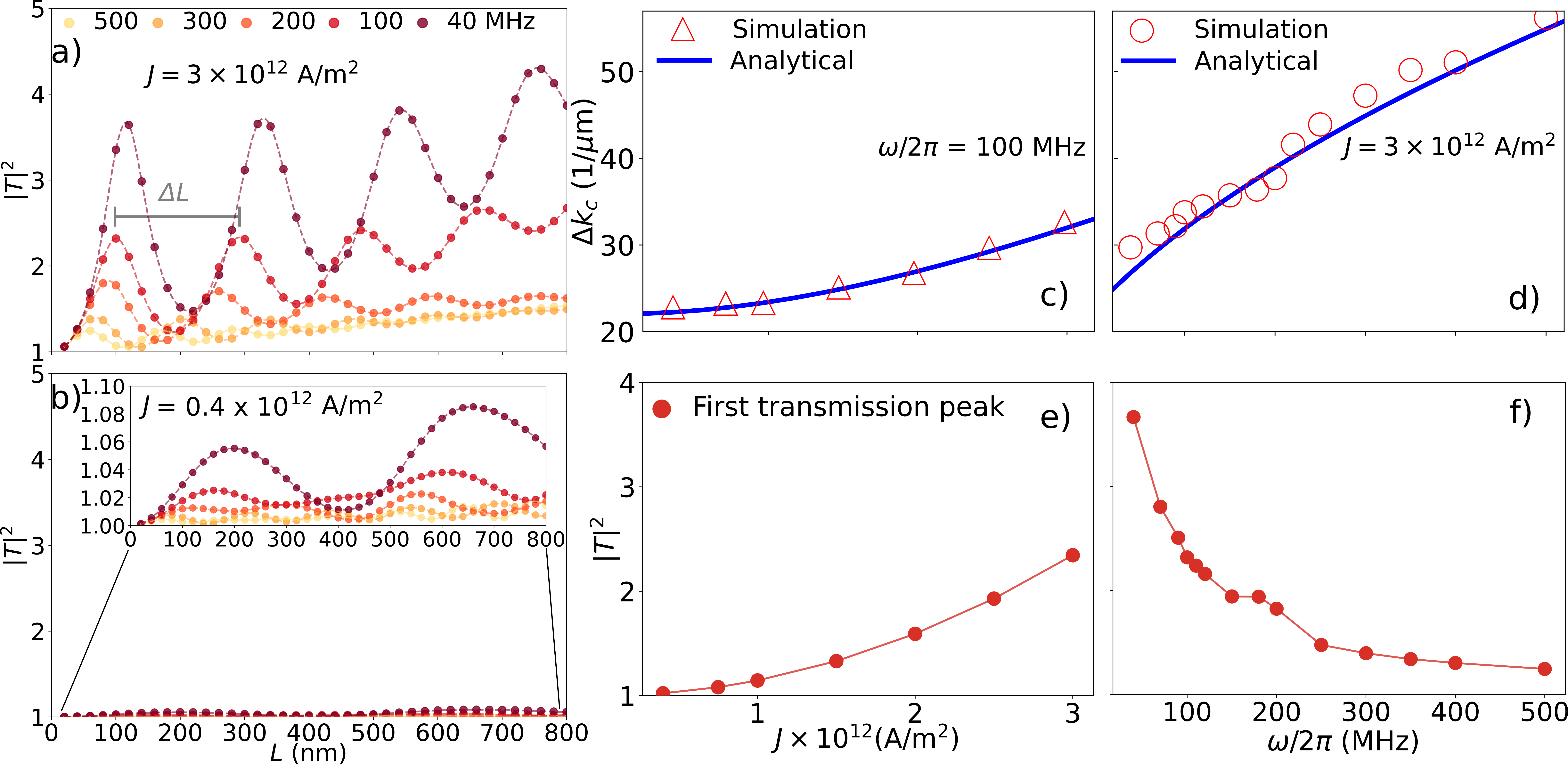}
\caption{\label{resonance_analysis} Spin-wave transmission characteristics and wavevector resonance analysis within the cavity. (a-b) Calculated spin-wave transmission as a function of cavity length for various excitation frequencies at a current density of (a) $J=3\times10^{12} \,\mathrm{A/m^2}$, \rai{with the characteristic resonance length $\Delta L$ indicated for the case of $100$~MHz}, and (b) $J=0.4\times10^{12} \,\mathrm{A/m^2}$. (c-d) Comparison of the wavevector variation $\Delta k$ obtained from numerical simulations (extracted via $\Delta L$, markers) and the analytical model ($\Delta k^\prime = k_2 - k_1$, solid blue lines) for (c) a fixed frequency of $100\,\mathrm{MHz}$ as a function of current density, and (d) a fixed current density of $J=3\times10^{12} \,\mathrm{A/m^2}$ as a function of frequency. (e-f) Amplitude of the first transmission peak \rai{as a function of (e) the current density, for $\omega/2\pi = 100\,\mathrm{MHz}$ and (f) the excitation frequency, with $J=3\times10^{12} \,\mathrm{A/m^2}$.}
}
\end{figure*}

Additionally, from Fig.~\ref{DHC_resonance_SW}(b), one can observe that the resonant behavior is not present for $J = 4\times10^{11}\,\mathrm{A/m^2} \approx 0.73J_c$, and it is considerably low for the cases where $J = 7.5\times10^{11}\,\mathrm{A/m^2} \approx 1.5J_c$ and $J = 1\times10^{12}\,\mathrm{A/m^2} \approx 2J_c$, even though $J > J_c$ inside the DHC. This highlights the need for interaction with negative-energy modes, which requires the emergence of multiple $k$ solutions that appear only when the SW frequency satisfies $\omega < \omega_\text{max}$, as discussed in the previous section. It is important to note that the transmission exhibits oscillatory behavior even for the lower current density considered in the simulations, \rmm{since reflected waves are also present inside the cavity, leading to standing waves but with much lower amplitudes.}

\rai{According to Refs.~\cite{SWLasing,nakayama2024resonant_christmas_eve}, the resonance condition is satisfied when the wave-vector solutions $k_3$ and $k_4$, shown in Fig.~\ref{Disp_rel_J1}(d), exist for the chosen SW frequency and satisfy $\Delta k = k_4 - k_3 = 2\pi n/L + \mathcal{O}(1/L^2)$, where $n$ is a positive integer. For our system, considering $J = 30\times10^{11}\,\mathrm{A/m^2}$, we obtain $\Delta k \approx 7~\upmu\mathrm{m}^{-1}$ by calculating the difference between the corresponding solutions of Eq.~(\ref{disp_STT}). However, the SW transmission obtained from the simulations yields $\Delta k_{\mathrm{sim}} \approx 33~\upmu\mathrm{m}^{-1}$. This value is much closer to the wave-vector difference $\Delta k^\prime = k_2 - k_1$, indicating that the observed resonant amplification in our system originates from the $k_1$ and $k_2$ modes rather than from the $k_3$ and $k_4$ solutions proposed in the literature. Furthermore, we investigated cavity lengths up to $L = 3000$~nm and found no evidence of resonances with the periodicity predicted by $\Delta k$ in the considered system. As will be shown in the following sections, however, such resonant solutions do exist under a different set of conditions.}

\rai{To understand the role of the excitation frequency in resonant amplification, Fig.~\ref{resonance_analysis}(a) shows the transmission amplitude as a function of cavity length for various excitation frequencies at a fixed current density of $J = 30\times10^{11}\,\mathrm{A/m^2}$. We observe that both the phase and the spacing between consecutive resonance peaks, $\Delta L$, vary significantly with frequency. In addition, the maximum transmission amplitude decreases as the excitation frequency increases. For $\omega/2\pi = 500\,\mathrm{MHz}$, the oscillatory behavior becomes strongly attenuated for larger cavity lengths. Notably, we do not observe a threshold frequency, $\omega_{\mathrm{max}}$, above which the resonant solutions vanish. Instead, the resonance amplitude decreases continuously with increasing frequency, further supporting the conclusion that the observed resonances originate from the $k_1$ and $k_2$ spin-wave modes rather than from the $k_3$ and $k_4$ modes. Finally, the magnitudes of the transmission peaks obtained here are consistent with those reported in Ref.~\cite{SWLasing}.}

Figure \ref{resonance_analysis}(b) shows the corresponding transmission for a subcritical current density, $J = 4\times10^{11}\,\mathrm{A/m^2}$. In this case, no significant resonant amplification is observed, as expected for $J<J_c$. However, a closer inspection of the data reveals residual oscillatory behavior, \rmm{as reflected waves are also present inside the cavity, leading to standing waves but with much lower amplitudes.}

\rai{To further characterize these observations, Figs.~\ref{resonance_analysis}(c,d) provide a direct quantitative comparison of the resonance period as a function of the applied current [Fig.~\ref{resonance_analysis}(c)] and the excitation frequency [Fig.~\ref{resonance_analysis}(d)]. The extracted wave-vector differences, $\Delta k_{\mathrm{sim}} = 2\pi/\Delta L$, where $\Delta L$ denotes the resonance period indicated in Fig.~\ref{resonance_analysis}(a), are shown as open symbols in Figs.~\ref{resonance_analysis}(c,d). The results demonstrate that the simulated values are in good agreement with $\Delta k^\prime = k_2 - k_1$ (blue solid lines), rather than with $\Delta k = k_4 - k_3$, further supporting our previous conclusion. Furthermore, the discrepancy arising from using $\Delta k$ instead of $\Delta k^\prime$ may explain the differences in the resonance behavior reported in recent studies of magnonic double-hole cavities~\cite{nakayama2024resonant_christmas_eve}.}

\rai{Fig.~\ref{resonance_analysis}(e) shows the amplitude of the first transmission peak as a function of current density. The data reveal an approximately parabolic increase in the transmission amplitude with increasing current density. The frequency dependence of the first transmission peak is shown in Fig.~\ref{resonance_analysis}(f). As the excitation frequency increases, the maximum transmission amplitude decreases smoothly, following an approximately continuous trend that mirrors the overall transmission decay observed in Fig.~\ref{resonance_analysis}(a).}
\begin{figure}[t!]
\includegraphics[width=\linewidth]{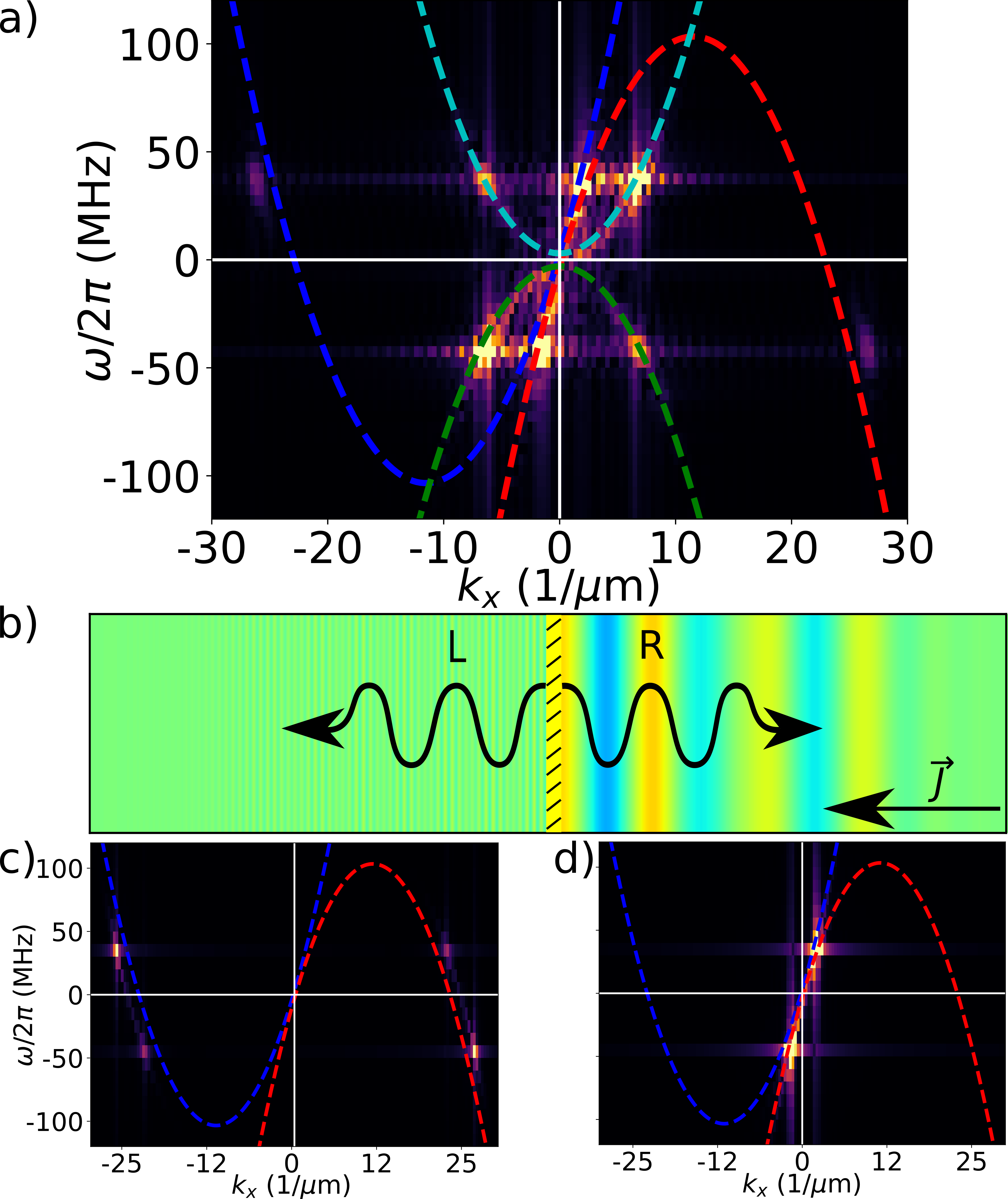}
\caption{\label{wavenumber_analysis} (a) Dispersion relation extracted from the sample containing the DHC [see the setup in Fig. \ref{DHC_resonance_SW}(a)] for a fixed frequency $\omega/2\pi = 40~\mathrm{MHz}$ and $L = 6000~\mathrm{nm}$. The cyan and green dashed parabolas represent the current-free solutions for the SW dispersion [Eq. (\ref{disp_STT})], while the red and blue dashed lines denote the analytical solutions for the region with $J = 3\times10^{12}~\mathrm{A/m^2}$. (b) Snapshot of the SW simulation for a sample without the DHC, with the antenna placed at the center of the sample and a current density of $J = 3\times10^{12}~\mathrm{A/m^2}$ applied uniformly along the entire stripe. The same magnetic parameters and antenna configuration as in (a) are used. (c,d) Dispersion relations for the system shown in (b), calculated on the (c) left-hand side and (d) right-hand side of the antenna, for $\omega/2\pi = 40~\mathrm{MHz}$. 
}
\end{figure}

\rai{To understand why no resonance involving the SW modes $k_3$ and $k_4$ is observed in the considered system, despite their clear presence in the dispersion relation [Fig.~\ref{Disp_rel_J1}(d)], we calculate the dispersion relation for a waveguide containing a DHC, whose geometry is illustrated in Fig.~\ref{DHC_resonance_SW}(a). An extended cavity length of $L = 6000\,\mathrm{nm}$ is used to enhance the spatial resolution inside the cavity. Figure~\ref{wavenumber_analysis}(a) shows the resulting dispersion relation for a fixed excitation frequency of $\omega/2\pi = 40\,\mathrm{MHz}$. The dashed lines represent the analytical solutions outside the cavity (cyan and green curves) and inside the cavity (blue and red curves). Notably, the SW modes belonging to the negative-energy branch, particularly the $k_4$ mode, are absent from the simulated dispersion, explaining why no significant resonant signal associated with these modes is observed in the considered waveguide. We attribute this behavior to the weak excitation of these modes during the reflection process inside the cavity. In particular, because the antenna is located outside the cavity, the $k_4$ mode, which has a negative group velocity, can only be generated as a reflected wave, resulting in an amplitude much smaller than that of the propagating modes.}

\rai{Fig.~\ref{wavenumber_analysis}(b) illustrates the setup used to calculate the dispersion relations shown in Fig.~\ref{Disp_rel_J1}. In this configuration, the current density is uniformly distributed along the entire length of the waveguide, and SWs are excited by an antenna that launches waves with both positive and negative group velocities. Figures~\ref{wavenumber_analysis}(c,d) show the dispersion relations extracted from (c) the left-hand side and (d) the right-hand side of the waveguide for a fixed excitation frequency of $\omega/2\pi = 40\,\mathrm{MHz}$. It can be seen that the $k_4$ mode appears only for SWs with negative group velocity [Fig.~\ref{wavenumber_analysis}(c)], and its amplitude is substantially smaller than that of the $k_1$ mode. Therefore, when the SW antenna is placed outside the cavity, the $k_4$ mode contributes only weakly to the reflected waves, which may suppress its contribution to the resonant amplification. Furthermore, exciting resonances involving these modes may require the SW source to be located inside the cavity, as is the case for the thermally excited SWs discussed later in this work.} 

\subsubsection{\label{sec:DHC_resonance_gaussian} DHC with a Gaussian-like current profile}

Our next step is to investigate whether SW resonant amplification persists under more realistic current profiles. As mentioned previously, a DHC for SWs can be achieved when the current density locally changes from $J < J_c$ to $J > J_c$ along the waveguide, for instance, in a constricted geometry where the current is crowded in the constriction. In previous works, the authors assumed a step-like current profile to describe the current density in such systems~\cite{SWLasing,nakayama2024resonant_christmas_eve}, achieving resonant amplification as we have demonstrated in the previous section. However, it is known that the current distribution in such constricted waveguides exhibits a finite gradient, with current components distributed in two dimensions, rather than a step-like transition in the vicinity of the DHC. In this section, we investigate the role of this current gradient in the realization of DHC-induced resonant amplification of SWs.

\begin{figure}[t!]
\includegraphics[width=\linewidth]{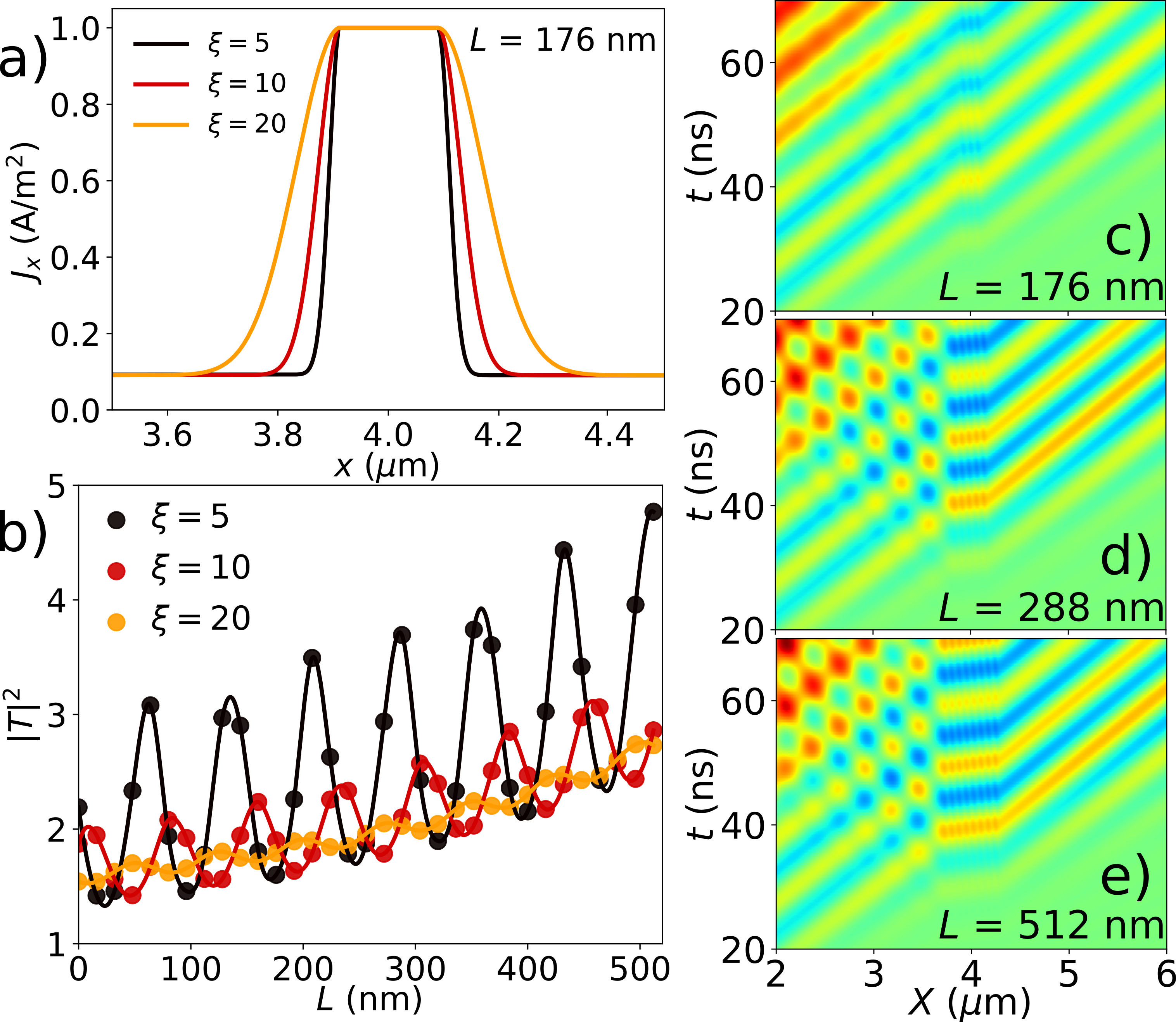}
\caption{\label{gaussian_resonance} (a) Normalized current distribution along the $x$-direction for the cavity length $L=176$~nm and different values of the parameter $\xi$ of the current distribution function. 
(b) SW transmission as a function of the cavity length $L$ for $J_{\max} = 10\times10^{12}\,\mathrm{A/m^2}$, considering different values of $\xi$. 
(c--e) Space--time diagrams for $\xi = 5$~nm: 
(c) off-resonance condition for $L = 176$~nm; 
(d) resonance condition for $L = 288$~nm (5 nodes); 
and (e) resonance condition for $L = 512$~nm (8 nodes inside the cavity). }
\end{figure}

Before modeling a fully two-dimensional layout, the isolated impact of a localized longitudinal current gradient is established using a 1D Gaussian-like profile. To this end, we performed simulations for the unconstricted waveguide, as in the previous section, but now assuming a Gaussian current-density profile with a plateau of constant value inside the DHC region; that is, 
$J(x) = J_{\max}$ when $x$ lies inside the DHC, and $J(x) = J_{\max} e^{-(x - x')^{2}/2\xi^{2}}$ when $x$ lies outside the DHC, where $x'$ denotes the position of the nearest DHC boundary. The current gradient at the DHC boundary is controlled by the parameter $\xi$. Figure~\ref{gaussian_resonance}(a) shows the current distribution along the $x$-direction for the cavity length $L=176$~nm and different values of the parameter $\xi$.

Figure~\ref{gaussian_resonance}(b) shows the SW transmission as a function of the cavity length $L$ for $J_{\max} = 10\times10^{12}\,\mathrm{A/m^2}$ and three selected values of the parameter $\xi$, representing large-gradient ($\xi = 5$ nm), moderate-gradient ($\xi = 10$ nm), and smooth-gradient ($\xi = 20$ nm) current profiles. The transmission curves reproduce the characteristic resonance peaks observed for the step-like current profile discussed in the previous section. However, a strong reduction in the amplitude of the transmission peaks is observed for smoother values of $\xi$, indicating that a sharp current variation is required for achieving DHC-induced resonant amplification of SWs. This may pose limitations for practical implementations, as very sharp or step-like current profiles are not straightforward to achieve in real samples.

Figs.~\ref{gaussian_resonance}(c--e) show the space--time diagrams for the case of $\xi = 5$~nm, for the off-resonance scenario at $L = 176$~nm [Fig.~\ref{gaussian_resonance}(c)], and the resonance scenarios at $L = 288$~nm [Fig.~\ref{gaussian_resonance}(d)] and $L = 512$~nm [Fig.~\ref{gaussian_resonance}(e)]. At the cavity center, standing waves with an integer number of nodes are clearly visible under resonance conditions. Moreover, the wave-number difference $\Delta k_{\text{sim}} = 85~\upmu\mathrm{m}^{-1}$ remains essentially constant for all three values of $\xi$, indicating that the resonance condition is governed by the current magnitude $J_{\max}$ and is not significantly affected by its spatial gradient.

\subsection{\label{MBH_real} Resonant amplification in a constricted waveguide}

To accurately model the formation of a DHC in a constricted waveguide and the interaction of SWs with the resulting non-uniform current distribution, we compute the two-dimensional current profile generated in the constriction using the Poisson solver implemented in our micromagnetic simulations, as described in Sec.~\ref{Methods}.

\begin{figure}[t!]
\includegraphics[width=0.99\linewidth]{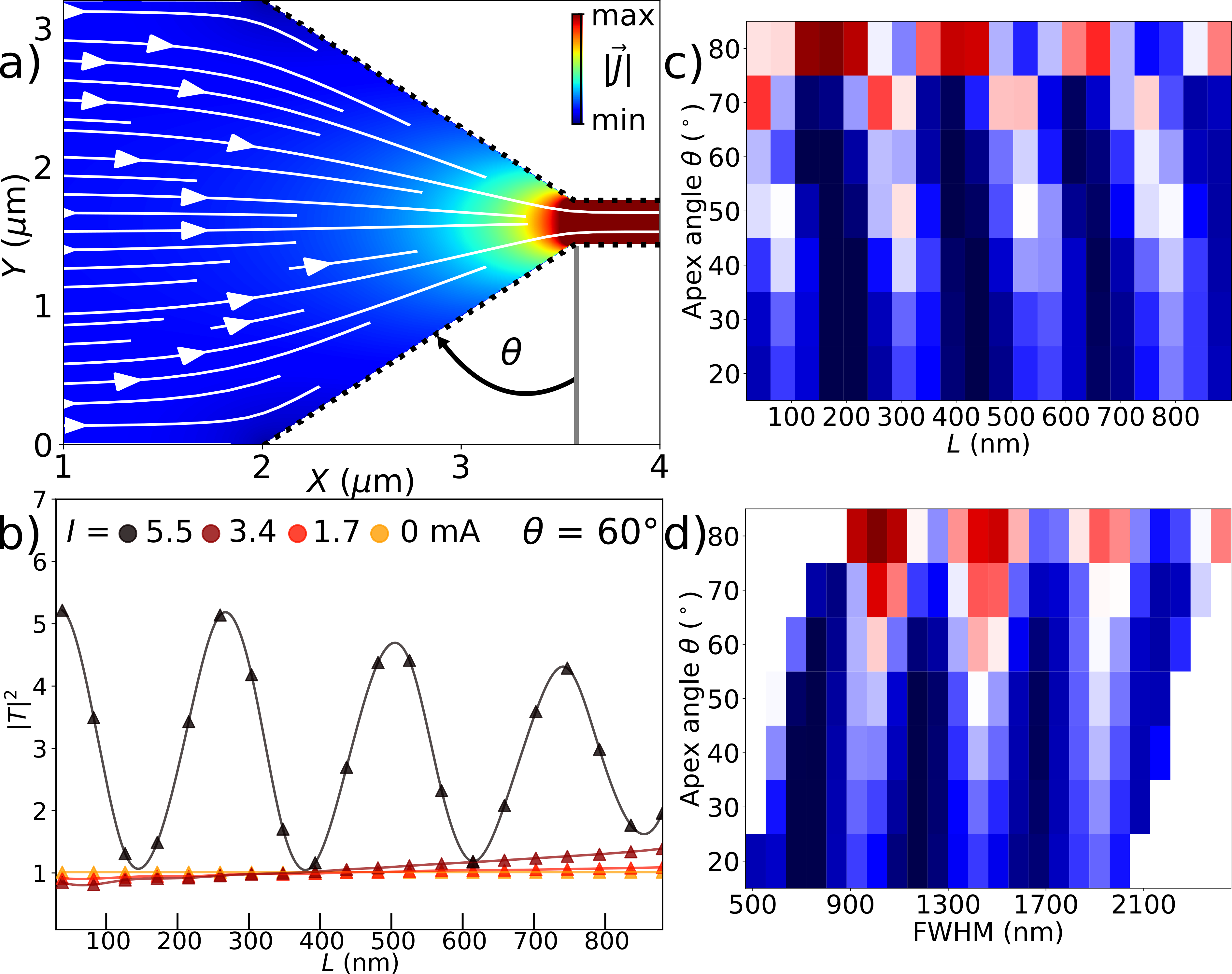}
\caption{\label{MBH_real_sample} (a) Illustration of the considered sample with an apex angle $\theta$. The current density for sample $\theta = 60^\circ$ is shown, with white arrows indicating the current direction. 
The transmission was therefore simulated for different total currents. (b) Spin-wave transmissions for sample $\theta = 60^\circ$ with different currents as a function of the cavity length $L$, for $I = 5.5$, $3.4$, $1.7$, and $0$~mA, corresponding to $J = 8.6~J_c$, $5.3~J_c$, $2.6~J_c$, and $0$, inside the cavity, respectively. (c) Overall normalized transmission signal for $I=5.5 \,\mathrm{mA}$ regarding different geometry angles considering each cavity size and (d) regarding the full width half maximum of their current distribution. The signal is normalized by the transmission amplitude of the $I = 0$ case for each sample.
}
\end{figure}

\rmm{For the simulations presented in this section, we consider several constricted samples with a total length of $8~\upmu\text{m}$, where the constriction geometry is modified by varying the apex angle $\theta$ of the constriction. The width of each sample changes from $W=3.2~\upmu\text{m}$ at the ends to $d_w=0.32~\upmu\text{m}$ at the constriction [see Fig.~\ref{Setup_config}]. The constriction apex angle is given by $\theta=\arctan(2l_t/(W-d_w))$. Figure~\ref{MBH_real_sample}(a) illustrates the sample geometry with calculated current density for the case of $\theta=60^\circ$. For a consistent comparison among the various geometries, the spin-wave transmission is analyzed as a function of the total applied current $I$ in each sample.}

Figure~\ref{MBH_real_sample}(b) presents the transmitted spin-wave amplitude as a function of the DHC length, $L$, \rmm{for excitation frequency of $\omega/2\pi=100$~MHz and apex angle $\theta = 60^\circ$}, and for representative values of the applied current, $I$. The signal is normalized to the zero-current transmission amplitude. \rmm{For the cases of $I = 1.7$ and $3.4 \,\mathrm{mA}$ ($J = 2.6J_c$ and $J = 5.3J_c$ inside the cavity)}, no resonance is observed across all samples, resulting in a transmission profile that closely tracks the current-free baseline, despite a small increase in the amplitude for larger cavity sizes. In contrast, transmission peaks characteristic of resonant behavior emerge at $I = 5.5 \,\mathrm{mA}$ ($J = 8.7J_c$ inside the cavity). This demonstrates resonant amplification within the constricted waveguides and confirms that a threshold current is necessary to form the DHC and sustain cavity resonances. 

To separate geometric from current-driven effects, SW propagation was also analyzed in the absence of current. In this case, no oscillations occur in the SW transmission when the cavity size $L$ is varied, confirming that the resonance arises exclusively from the current-induced DHC.
To further elucidate the role of the constriction geometry on these resonant modes, the overall normalized transmission signal was systematically investigated across a broad range of apex angles $\theta$, as depicted in the heatmap of Fig. \ref{MBH_real_sample}(c). The spatial distribution of the resonances is highly sensitive to this geometric gradient; as the apex angle $\theta$ increases, the periodic high-amplitude resonant bands shift systematically toward shorter cavity lengths $L$, confirming that standing-wave formation is strictly governed by local \rmm{longitudinal} confinement. Beyond this spatial shifting, the transmission amplitude itself exhibits a pronounced geometric dependence.

\rmm{This geometric dependence is particularly intriguing when compared with the idealized one-dimensional models of the DHC. For a 1D Gaussian-like current profile, as shown in the previous section, simulations suggest that a sharp spatial variation in current (i.e., small values of $\xi$) maximizes resonant amplification. Therefore, one might expect that the sharper constrictions (e.g., $\theta = 20^\circ$), which most closely mimic the step-like current scenario, would yield the highest transmission. However, comparison among the constricted geometries with apex angles between $30^\circ$ and $60^\circ$ reveals the opposite behavior: the maximum transmission actually occurs for the smoother current gradient (i.e., $\theta = 60^\circ$), rather than for the sharper constrictions. This discrepancy may be related to geometrical effects that are absent in the idealized 1D model, such as SW reflections at the constriction entrance, which are stronger for sharper constrictions than for smoother ones. These reflections can reduce the amount of the wave entering the DHC region and consequently affect the transmission amplitudes.}

\rmm{Furthermore, as $\theta$ increases, the resonance peaks in Fig.~\ref{MBH_real_sample}(c) shift toward lower values of $L$. This occurs because a more gradual taper (larger $\theta$) causes the current density to exceed $J_c$ farther away from the central flat region, effectively increasing the active cavity size, $L_{\mathrm{eff}} > L$. Consequently, to satisfy the fixed phase-matching resonance condition, the nominal length $L$ must decrease. This interpretation is corroborated by analyzing the transmission as a function of the full width at half maximum (FWHM) of the current distribution, as presented in Fig.~\ref{MBH_real_sample}(d). While sharp constrictions successfully generate a highly concentrated current spike, they inherently possess a narrower FWHM. Conversely, more gradual tapers distribute the elevated current density over a significantly wider spatial region. Figure~\ref{MBH_real_sample}(d) reveals that the regions of maximum resonant transmission consistently align with the FWHM values corresponding to the different apex angles.}

\rmm{These results demonstrate that resonant spin-wave amplification is robust in DHCs with realistic constricted geometries, and that optimal device performance requires careful engineering of the apex angle to mitigate SW reflections.}

\subsection{\label{subsec:Kz}Effect of uniaxial anisotropy}

\rai{To evaluate the influence of magneto-crystalline anisotropy in the resonant amplification, we simulated SW propagation using a setup similar to that in Sec.~\ref{sec.step-like}, i.e. a DHC with a step-like current profile, while introducing an out-of-plane uniaxial anisotropy $K_z$ across the entire sample. This investigation is motivated by previous reports demonstrating that tuning the spin-wave polarization from circular to elliptical is a critical prerequisite for achieving efficient resonant amplification~\cite{SWLasing}.}

\rai{
In the presence of uniaxial anisotropy, the background magnetization can deviate from the direction of the bias field. The equilibrium state can be determined analytically by considering the rotation of the magnetization within the $y$--$z$ plane. In this case, the magnetization vector can be parameterized as $\vec{m} = (0, \sin\theta, \cos\theta)$, where $\theta$ is the angle measured from the $z$-axis. Minimizing the system energy [Eq.~(\ref{eq.Energy})] with respect to $\theta$ yields the equilibrium condition $\cos\theta \left( \frac{2K_z}{M_s}\sin\theta - B_y \right) = 0$. This equation gives rise to two distinct equilibrium regimes. The first solution, $\cos\theta = 0$ (i.e., $\theta = \pm\pi/2$), corresponds to the magnetization being fully aligned with the external field along the $y$-direction. The second solution is obtained by setting the term in parentheses equal to zero, resulting in $\sin\theta = B_y M_s/(2K_z)$. This solution exists only when $K_z \geq K_c = B_y M_s/2$, which ensures that $\sin\theta \leq 1$. For the magnetic parameters considered in this work, the critical anisotropy separating these two regimes is $K_c = 37.5~\mathrm{J/m^3}$.}

\begin{figure}[t!]
\includegraphics[width=\linewidth]{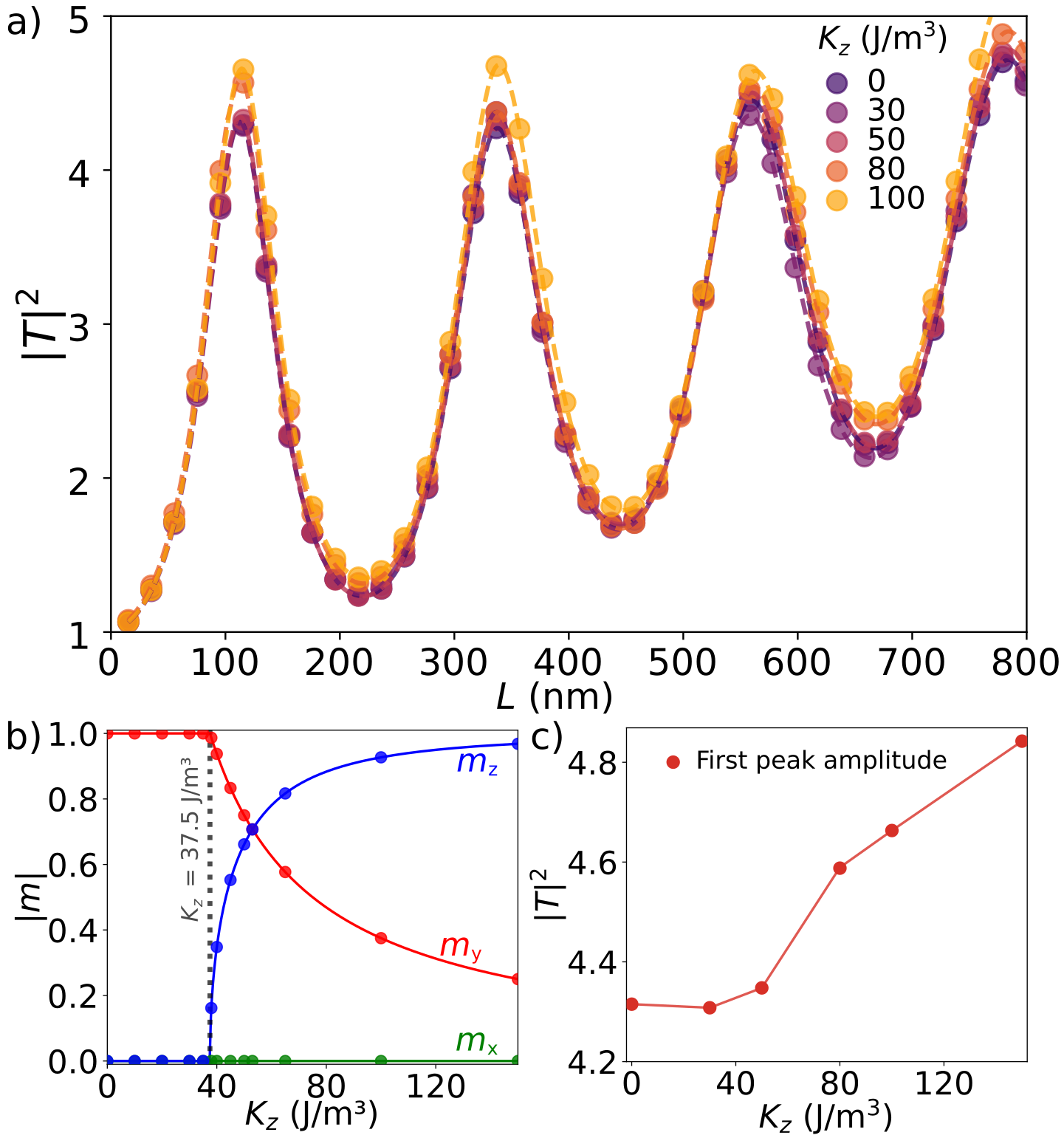}
\caption{\label{Effect_anis} Influence of out-of-plane uniaxial anisotropy on SW transmission. (a) SW transmission amplitude as a function of the cavity length $L$ for selected values of $K_z$. (b) Equilibrium magnetization components $m_x$, $m_y$, and $m_z$ as a function of $K_z$. Solid lines represent the analytical results, while circles denote micromagnetic simulation data. The critical anisotropy $K_c$, below which the magnetization is fully aligned with the external field, is indicated by the gray dashed line. (c) Transmission amplitude of the first resonance peak as a function of $K_z$.
}
\end{figure}

\rai{
Figure~\ref{Effect_anis}(a) shows the transmission amplitude as a function of the cavity length for different values of $K_z$, while Fig.~\ref{Effect_anis}(b) presents the equilibrium magnetization components as a function of $K_z$. The solid lines correspond to the analytical solutions, and dots to the simulation data. For $K_z < 37.5,\mathrm{J/m^3}$, the magnetization remains entirely in-plane ($m_y = 1$), as expected. From Fig.~\ref{Effect_anis}(a), one can notice that $K_z$ has only a negligible effect on the transmission profile, even for anisotropy values above the critical value ($K_z > K_c$). This is evidenced by the small variation in the amplitude of the first transmission peak, as quantified in Fig.~\ref{Effect_anis}(c). Since varying $K_z$ within the low-anisotropy regime produces only negligible changes in the resonant amplification, the effect of uniaxial anisotropy is neglected in the subsequent analysis.}

\subsection{\label{subsec:DMI}Reducing critical current and increasing amplification by DMI}

While magnonic DHC are discussed here in the context of realistic current distributions dictated by sample geometry, the high current densities required in such scenarios pose substantial challenges for experimental implementation. In particular, the associated Joule heating is expected to deteriorate spin-wave coherence. To address this limitation, we propose employing iDMI to facilitate the formation of DHCs with reduced current densities.

As known in the literature \cite{mulkers2018tunable,DMI_SW,SWLasing}, DMI introduces an additional source of asymmetry in the spin-wave dispersion, analogous to the effect of the current density, resulting in an extra term that depends linearly on $k$. By including the interfacial DMI energy
\[
\mathcal{E}_{\mathrm{DMI}} = D \left[ m_z (\nabla \cdot \mathbf{m}) - (\mathbf{m} \cdot \nabla) m_z \right],
\]
where $D$ denotes the DMI strength, in the Hamiltonian of Eq.~\eqref{eq.Energy}, one obtains an expression for the SW dispersion similar to Eq.~(\ref{disp_STT}), but with the current-induced contribution $v_s$ replaced by the background velocity associated with DMI, $v_{\mathrm{DMI}}$, which can be written as \cite{DMI_SW}
\begin{equation}
    v_{\mathrm{DMI}} = \frac{2\gamma D}{M_S}.
    \label{vDMI}
\end{equation}
Therefore, one can expect DMI to produce an effect similar to that of the current density in the formation of a DHC, such that including this interaction enables the creation of a DHC at considerably reduced current densities. Because DMI effectively compensates for a portion of the required current, the two mechanisms can be superimposed. For small values of the non-adiabatic factor $\beta$, the total effective background velocity $v_{bg}$ is expressed as:
\begin{equation}
    v_{\mathrm{bg}} = \frac{PJ\mu_B}{eM_s} +\frac{2\gamma D}{M_S}. \label{bgvelocity}
\end{equation}

\begin{figure}[t]
\includegraphics[width=\linewidth]{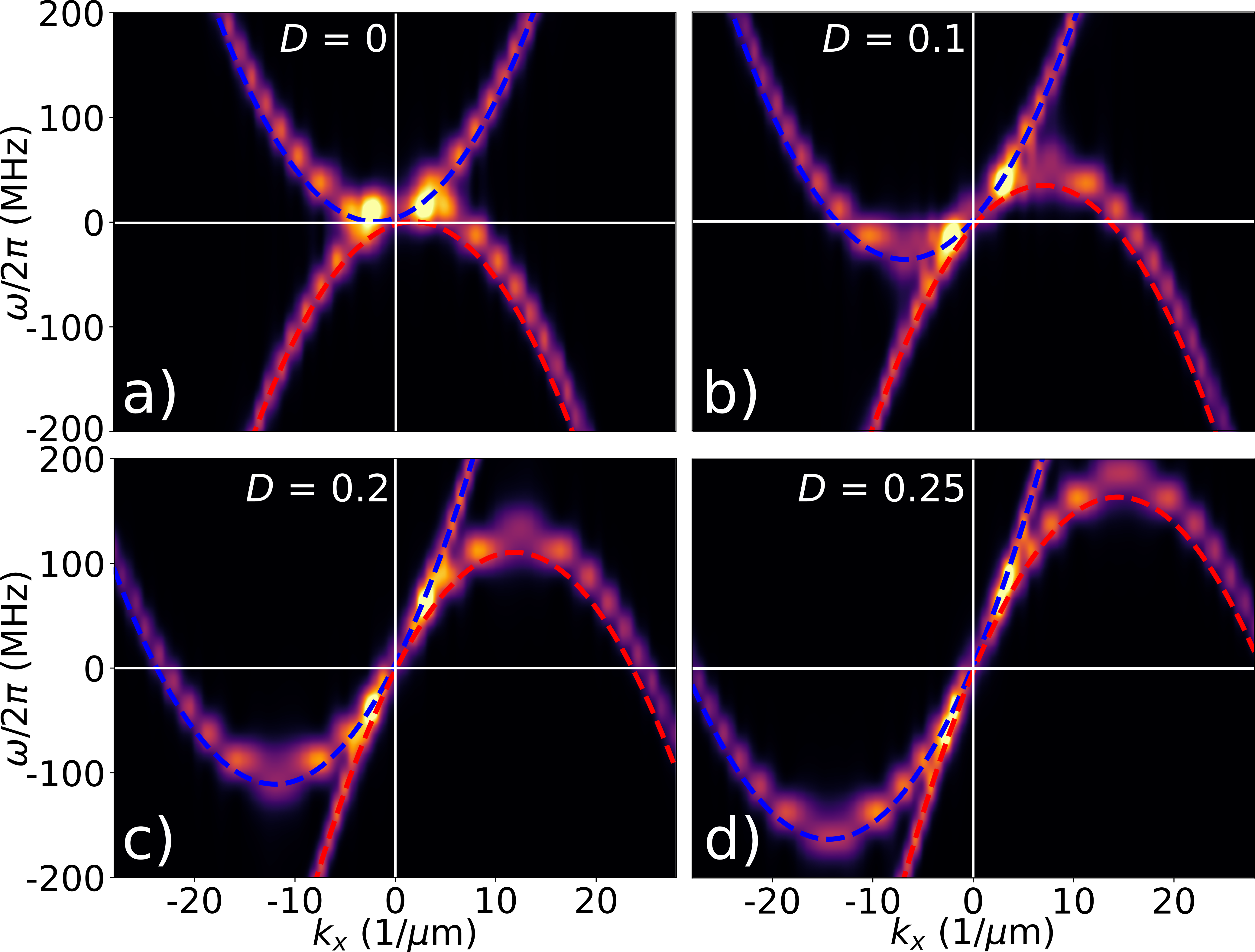}
\caption{\label{Disp_rel_DMI} Dispersion relation for the unconstricted permalloy strip for 
$J = 5 \times 10^{11}\,\text{A/m}^2 \approx J_c$ and the interfacial DMI:
(a) $D = 0$, (b) $D = 0.1$, (c) $D = 0.2$, and (d) $D = 0.25$ mJ/m$^2$. 
The red and blue dashed lines correspond to the $+\omega$ and $-\omega$ solutions 
obtained by combining Eqs.~(\ref{disp_STT}) and (\ref{bgvelocity}).
 }
\end{figure}

Figure~\ref{Disp_rel_DMI} shows the SW dispersion relations calculated for a fixed current $J = 5 \times 10^{11}\,\text{A/m}^2 \approx J_c$ and different values of the DMI strength. A comparison of Fig. \ref{Disp_rel_DMI} with Fig. \ref{Disp_rel_J1} reveals that increasing the DMI yields spin-wave dispersions equivalent to those driven by higher current densities. Moreover, the red and blue dashed lines in Fig.~\ref{Disp_rel_DMI} correspond to the $+\omega$ and $-\omega$ solutions obtained by combining Eqs.~(\ref{disp_STT}) and (\ref{bgvelocity}), thus demonstrating that the calculated dispersions are in good agreement with the analytical predictions.

The magnitude of the DMI that mimics a specific value of the current density can be obtained by comparing the induced background velocities in Eq.~\eqref{bgvelocity}, from which one finds $D = \frac{\mu_B P J}{2 e \gamma}$. Therefore, for the parameters considered in this work, a current density of, e.g., $J = 30 \times 10^{11}\,\text{A/m}^2 \approx 6J_c$ can be mimicked by $D \approx 0.23~\text{mJ/m}^2$, which is a feasible value for realistic samples \cite{DMI_SW,DMI_interfacial}. This result is particularly noteworthy, as even a modest DMI strength can reproduce the effect of large current densities. Crucially, the magnitude of the interfacial DMI must be carefully optimized, as excessively large DMI values can stabilize chiral textures (\textit{e.g.}, spin spirals or domain walls), thereby degrading coherent spin-wave propagation.

\begin{figure}[t!]
\includegraphics[width=\linewidth]{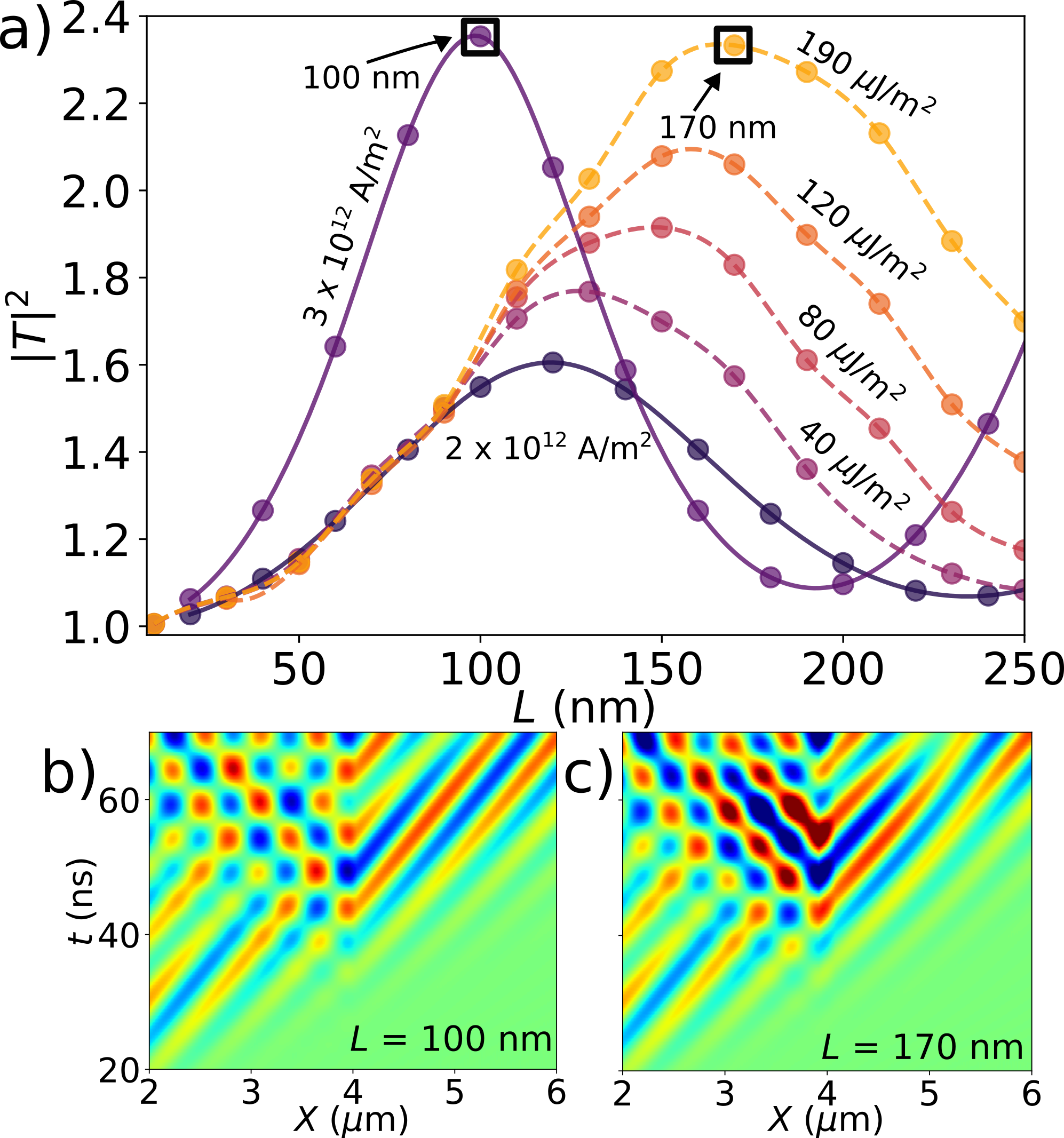}
\caption{\label{DHC_resonance_SW+DMI} (a) SW transmission in a rectangular stripe for $J = 20\times10^{11}$ and $30\times10^{11} \,\mathrm{A/m^2}$(solid lines, cf. Fig.~\ref{DHC_resonance_SW}(b)) and different interfacial DMI inside the cavity for $J = 20\times10^{11} \,\mathrm{A/m^2}$. Space-time diagrams for the transmission peak for (b) $J = 30\times10^{11} \,\mathrm{A/m^2}$ at $L = 100 \,\mathrm{nm}$, and for (c) $J = 20\times10^{11} \,\mathrm{A/m^2}$ and $D = 190 \,\mathrm{\upmu J/m^2}$ at $L = 170 \,\mathrm{nm}$. }
\end{figure}

To demonstrate the effect of DMI on DHC-induced SW amplification, we considered the same simulation setup as in Sec.~\ref{sec.step-like}, i.e., a DHC with a step-like current profile, but with a uniform DMI applied within the cavity region. The simulation was performed in three distinct stages. First, the system was allowed to reach a relaxed ground state with the DMI applied. Second, the current was turned on inside the cavity for 100 ns. Finally, once this steady state was established, the antenna was activated to excite and transmit the spin waves.

Figure~\ref{DHC_resonance_SW+DMI}~(a) presents the spin-wave transmission as a function of cavity length, comparing purely current-driven scenarios with cases in which the DHC is supplemented by DMI. We observe that, as the DMI strength increases, the amplification amplitude increases significantly. In particular, a current density of $J = 20\times10^{11} ~\mathrm{A/m^2}$ combined with a DMI strength of $D = 190 ~\mathrm{\upmu J/m^2}$ yields a peak transmission comparable to that obtained at the much higher current density of $J = 30\times10^{11} ~\mathrm{A/m^2}$ in the absence of DMI. Moreover, the position of the transmission peak shifts to $L = 170$~nm, compared with $L = 100$~nm for the zero-DMI case at $J = 30\times10^{11}~\mathrm{A/m^2}$.
These results confirm that DMI cooperates with the applied current to enhance resonant amplification, thereby increasing spin-wave transmission while requiring significantly lower current densities.

The space-time diagrams further corroborate this DMI-assisted amplification mechanism. Figure~\ref{DHC_resonance_SW+DMI}(b) shows the purely current-driven resonance at $J = 30\times10^{11}~\mathrm{A/m^2}$ and $L = 100$~nm, exhibiting the expected standing-wave pattern and high transmission amplitude. In contrast, Fig.~\ref{DHC_resonance_SW+DMI}(c) illustrates the DMI-assisted resonance at $J = 20\times10^{11}\,\mathrm{A/m^2}$, with $D = 190\,\mathrm{\upmu J/m^2}$ and $L = 170$~nm, where similar transmission amplitude is obtained. Thus, incorporating DMI provides a promising and practical strategy for lowering the threshold current required for resonant amplification, thereby reducing Joule heating.

\subsection{\label{subsec:SW_lasing} Spin-wave lasing}

Together with the resonant amplification observed in the DHC system, these results suggest the possibility of realizing a spin-wave laser, as proposed in Ref.~\cite{SWLasing}, in which thermally induced SWs generated within a heated constriction or cavity can undergo resonant amplification and be emitted with enhanced amplitudes at selected frequencies satisfying the resonance conditions. To explore this scenario, we introduce thermal excitation into the DHC simulations. Specifically, a stochastic thermal field $H_{\mathrm{th}}$ is included in the calculations, as implemented in the Mumax$^3$ package~\cite{Mumax_original2014}, such that the effective field is given by $H_{\mathrm{eff}} = -\delta E/\delta \vec{M} + H_{\mathrm{th}}$. In this section, we consider the same simulation setup as in Sec.~\ref{sec.step-like}, i.e., a DHC with a step-like current profile, but with a fixed nonzero temperature inside the cavity region. The simulations are performed for a permalloy strip with a length of $32~\mathrm{\mu m}$, a nominal width of $w=800~\mathrm{nm}$, and a thickness of $20~\mathrm{nm}$, with the center of the DHC located at $x=16~\mu\mathrm{m}$. This system shares key physical similarities with a constricted geometry subjected to an applied current, where the constricted region naturally leads to a higher current density and, consequently, an increased local temperature, thereby capturing the behavior described by our simplified model.

\begin{figure}[t!]
\includegraphics[width=\linewidth]{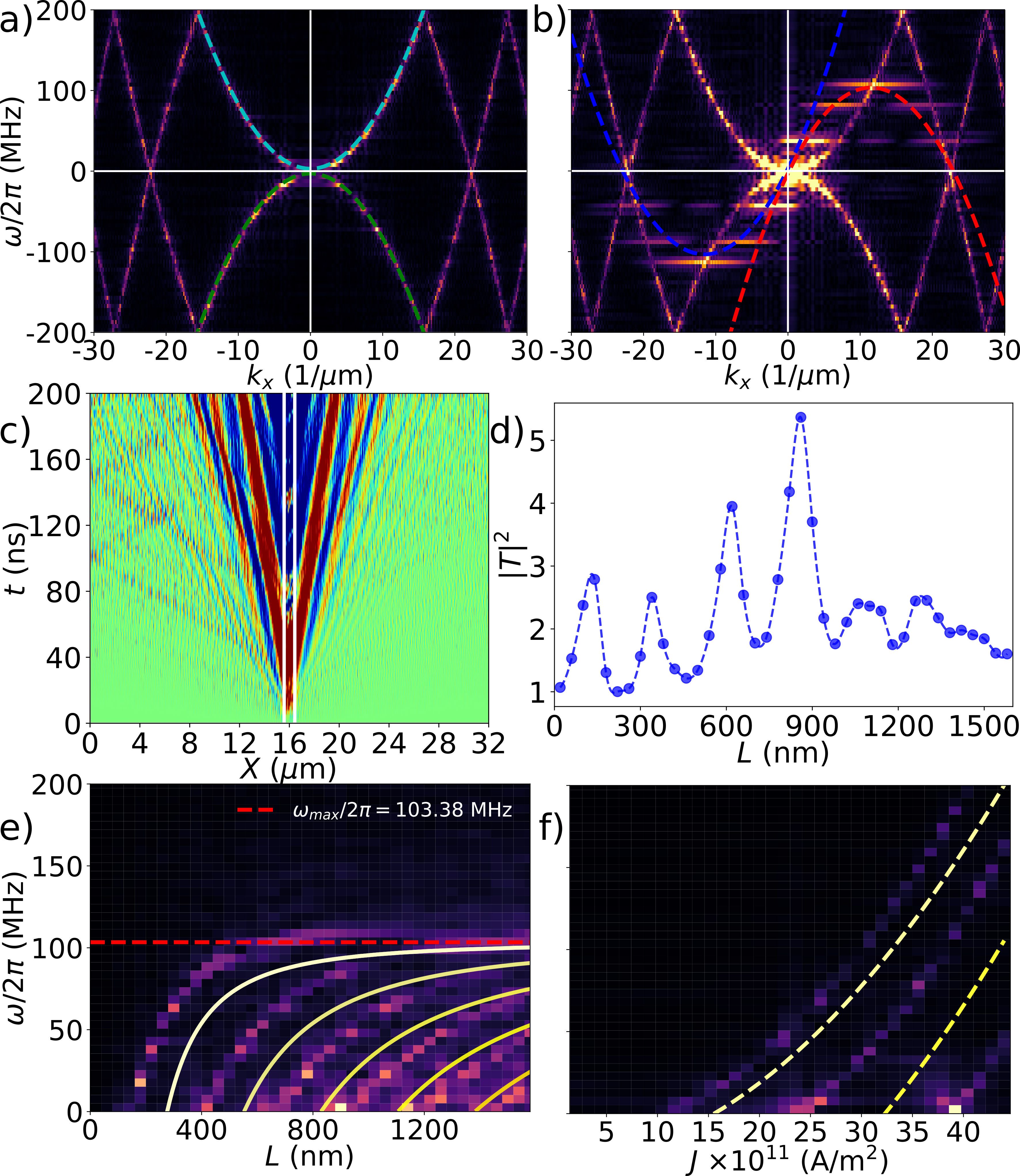}
\caption{\label{SW_lasing_prototype} (a,b) Dispersion relations under thermal excitation for a cavity length of $L = 860~\mathrm{nm}$ for (a) $J = 0$ and (b) $J = 3 \times 10^{12}~\mathrm{A/m^2}$. The cyan and green dashed parabolas in (a) represent the current-free solutions for the SW dispersion [Eq. (\ref{disp_STT})], while the red and blue dashed lines in (b) denote the analytical solutions for $J = 3\times10^{12}~\mathrm{A/m^2}$. (c) Space-time diagram for $L = 860~\mathrm{nm}$, where vertical white lines delineate the cavity boundaries, illustrating spin-wave lasing. (d) Normalized SW emission intensity $|T|^2$ evaluated at $x = 12~\mathrm{\upmu m}$ as a function of cavity length $L$, normalized to the reference case at $J = 0$. (e,f) SW dispersion as a function of $L$ for $J = 3\times10^{12}~\mathrm{A/m^2}$ in (e) and as a function of $J$ for $L = 500~\mathrm{nm}$ in (f). The solid yellow lines correspond to the analytical solutions obtained from Eq.~\ref{frequency_vs_current_cavity} for different values of $n$.}
\end{figure}

Figures \ref{SW_lasing_prototype}(a) and \ref{SW_lasing_prototype}(b) depict the simulated dispersion relations for the considered system in the cases of $J = 0$ (a) and $J = 3 \times 10^{12}\,\mathrm{A/m}^2$ (b) inside the cavity. The temperature inside the cavity is set to $50$~K in both cases. The analytical solutions for the dispersion relations at the respective current densities are shown as dashed lines, demonstrating excellent agreement with the simulation data. In particular, for $J = 3 \times 10^{12}\,\mathrm{A/m}^2$ [Fig. \ref{SW_lasing_prototype}(b)], the SW dispersion is characterized by the amplification of specific frequency excitations, with transmission peaks occurring at selected frequencies along the predicted SW dispersion relation. These selected frequencies indicate that, in this scenario, the thermally induced SWs undergo resonant amplification.

Fig. \ref{SW_lasing_prototype}(c) shows the space-time diagram for the case of $L=860$~mm, where the magnetization profile was summed along the sample width ($y$ direction) for better visualization. Notably, coherent SW emission can be observed from the center of the waveguide, where the DHC is located, propagating in both the $-\hat{x}$ and $+\hat{x}$ directions, thus demonstrating the spin-wave lasing phenomenon in the considered system.

The SW emission as a function of cavity length $L$ is shown in Fig.~\ref{SW_lasing_prototype}(d). Periodic emission peaks are observed, which is characteristic of resonant amplification. According to Ref.~\cite{SWLasing}, a SW laser can be realized when the resonance condition is satisfied for the wave-vector solutions $k_3$ and $k_4$, shown in Fig.~\ref{Disp_rel_J1}(d), at a given SW frequency. These solutions satisfy $\Delta k = k_4-k_3 = 2\pi n/L + \mathcal{O}(1/L^2)$, where $n$ is a positive integer. To identify which SW modes satisfy the resonance conditions in our system, we analyze the frequencies of the emitted SW modes as a function of the cavity length and applied current. Considering the wave-vector mismatch $\Delta k=k_4-k_3$ obtained from Eq.~\ref{disp_STT}, we can write
\begin{equation}
    \frac{2\gamma A}{M_S}\Delta k = \sqrt{(\frac{P\mu_BJ}{eM_S})^2+4(\frac{2\gamma A}{M_S})(\omega_0-2\pi f)}, \label{critical_values}
\end{equation} 
where $f$ is the excitation frequency of the system. By imposing the cavity resonance condition $\Delta k=2\pi n/L$, Eq.~\ref{critical_values} can be rearranged in terms of $f$ as
\begin{equation}
    2\pi f(J,L) = \frac{P^2\mu_B^2J^2}{8e^2M_S\gamma A} - \frac{2\gamma A\pi^2n^2}{M_SL^2} +\gamma B_y. \label{frequency_vs_current_cavity}
\end{equation}
Equation~\ref{frequency_vs_current_cavity} describes the frequencies of the emitted SW modes, corresponding to different excitation modes $n$, as a function of the current density $J$ and cavity length $L$.

Figures~\ref{SW_lasing_prototype}(e) and \ref{SW_lasing_prototype}(f) show the SW dispersion as a function of $L$ for $J=3\times10^{12}~\mathrm{A/m^2}$ [Fig.~\ref{SW_lasing_prototype}(e)] and as a function of $J$ for $L=500$~nm [Fig.~\ref{SW_lasing_prototype}(f)]. The solid yellow lines correspond to the analytical solutions obtained from Eq.~\ref{frequency_vs_current_cavity}. Despite the small shift between the numerical results and the analytical solutions, the observed modes and their dependence on $L$ and $J$ are in good agreement with the analytical calculations. Moreover, the critical frequency $\omega_{\mathrm{max}}$, above which the SW solutions $k_3$ and $k_4$ cease to exist, is shown for $J=3\times10^{12}~\mathrm{A/m^2}$ as a red dashed line in Fig.~\ref{SW_lasing_prototype}(e). Note that no emitted modes are observed above this limit, further corroborating that the resonant amplification is governed by these SW modes.       

While these explicit simulations are based on an idealized model, the presented results can be extended to realistic waveguides with physical constrictions. In a practical device, although a localized high-current-density region remains present within the constriction, which is essential for defining the DHC, the baseline current density in the rest of the waveguide is not zero, as assumed in the present model, but rather remains below the critical current density ($J<J_c$). This background current introduces a controllable breaking of the spatial inversion symmetry for SW propagation. Consequently, the finite current density outside the constriction breaks the spatial symmetry along the waveguide, leading to asymmetric SW emission and imparting a preferential propagation direction to the resulting lasing modes, rather than the symmetric emission observed in Fig.~\ref{SW_lasing_prototype}(c). This effect can ultimately enable SW lasing with unidirectional emission.

\section{\label{sec:End}Conclusion}

In conclusion, we demonstrated and characterized the resonant amplification of spin waves in double-hole cavities (DHCs) embedded in current-carrying constricted magnetic waveguides by combining analytical models with micromagnetic simulations. For the idealized step-like current profile, we reproduced the predicted resonant amplification as a function of current density and SW frequency and identified the spin-wave modes responsible for the resonances. We find that, when SWs are generated outside the DHC, the resonant modes differ from those previously reported in the literature, providing new insights into the origin of the resonant response.

We further considered realistic two-dimensional current distributions obtained using a Poisson solver and demonstrated that the resonant amplification persists in realistic constricted geometries. We showed that the resonance can be tuned through both geometrical and magnetic parameters. In particular, the interfacial Dzyaloshinskii--Moriya interaction (iDMI) can effectively mimic the role of the applied current, reducing the threshold current and associated Joule heating and thus facilitating the experimental realization of magnonic gravity analogues.

Finally, by including thermal fluctuations, we demonstrated spin-wave lasing in the DHC. Thermally generated spin waves are selectively amplified and coherently emitted at well-defined resonance frequencies. In this case, the resonant modes are consistent with those predicted in previous theoretical studies~\cite{SWLasing}. Overall, our results establish the DHC as a tunable platform for resonant spin-wave amplification and coherent magnon generation, with potential applications in low-power and nonreciprocal magnonic devices.

\section*{Acknowledgement}
This work was supported by the Research Foundation-Flanders (FWO) and the Special Research Funds of the University of Antwerp (BOF-UA), as well as the Brazilian agencies CNPq (INCT MQA Proc. 408766/2024-7), FAPESP, and CAPES. The computational resources used in this work were provided by the VSC (Flemish Supercomputer Center), funded by Research Foundation-Flanders (FWO) and the Flemish Government -- department EWI.
\clearpage
\bibliographystyle{ieeetr}
\bibliography{references}
\clearpage

\appendix
\end{document}